\documentclass[screen, acmsmall]{acmart}
\usepackage{dirtytalk}
\usepackage[nolist,nohyperlinks]{acronym}
\usepackage{multirow}
\usepackage[draft]{minted}

\setminted[c]{%
    breaklines,%
    fontsize=\footnotesize,
    }%
\makeatletter
\newenvironment{tabminted}{%
    \let\FV@ListVSpace\relax  
    \minted
}{%
    \endminted
    \unskip   
    \aftergroup\@tabmintedend
}
\newcommand*{\tabminted@finalstrut}[1]{%
    \ifdim\prevdepth>0pt
        \ifdim\dp#1>\prevdepth
            \vskip\dimexpr(\dp#1)-\prevdepth\relax
        \fi
    \else
        \vskip\dimexpr(\dp#1)\relax
    \fi
}
\newcommand*{\@tabmintedend}{%
    \let\@finalstrut\tabminted@finalstrut
}
\makeatother
\usepackage{lscape} 

\AtBeginDocument{%
  \providecommand\BibTeX{{%
    \normalfont B\kern-0.5em{\scshape i\kern-0.25em b}\kern-0.8em\TeX}}}

\setcopyright{acmcopyright}
\acmJournal{TOCHI}
\acmYear{2022} \acmVolume{29} \acmNumber{2} \acmArticle{17} \acmMonth{1} \acmPrice{15.00}\acmDOI{10.1145/3489465}

\newcommand{\sys}{Themisto}
\newcommand{\inlinequote}[1]{\say{\textit{#1}}}
\newcommand\revision[1]{{\color{black} #1}}

\begin{acronym}
    \acro{URL}{Uniform Resource Locator}
    \acro{AI}{Artificial Intelligence}
    \acro{ML}{Machine Learning}
    \acro{DS}{Data Science}
    \acro{GNN}{Graph Neural Network}
    \acro{HCI}{Human-Computer Interaction}
    \acro{CSED}{Computer Science Education}
    \acro{CSCW}{Computer-Supported Cooperative Work}
    \acro{UI}{User Interface}
    \acro{GUI}{Graphical User Interface}
    \acro{AIOP}{AI Operator}
    \acro{AST}{Abstract Syntax Tree}
    \acro{NLP}{Natural Language Processing}
    \acro{NMT}{Neural Machine Translation}
    \acro{API}{Application Programming Interface}
    \acro{HTTP}{Hypertext Transfer Protocol}
\end{acronym}

\begin{document}



\title[Human-Centered AI System to Assist Data Science Code Documentation in Computational Notebooks]{\revision{Documentation Matters: Human-Centered AI System to Assist Data Science Code Documentation in Computational Notebooks}}

\author{April Yi Wang}
\authornote{Both authors contributed equally to this research.}
\email{aprilww@umich.edu}
\affiliation{
\institution{University of Michigan}
\country{USA}
}
\author{Dakuo Wang}
\authornotemark[1]
\email{dakuo.wang@ibm.com}
\affiliation{
\institution{IBM Research}
\country{USA}
}
\author{Jaimie Drozdal}
\email{drozdj3@rpi.edu}
\affiliation{
\institution{Rensselaer Polytechnic Institute}
\country{USA}
}
\author{Michael Muller}
\email{michael_muller@us.ibm.com}
\affiliation{
\institution{IBM Research}
\country{USA}
}
\author{Soya Park}
\email{soya@mit.edu}
\affiliation{
\institution{MIT CSAIL}
\country{USA}
}
\author{Justin D. Weisz}
\email{jweisz@us.ibm.com}
\affiliation{
\institution{IBM Research}
\country{USA}
}
\author{Xuye Liu}
\email{liux27@rpi.edu}
\affiliation{
\institution{Rensselaer Polytechnic Institute}
\country{USA}
}
\author{Lingfei Wu}
\email{wuli@us.ibm.com}
\affiliation{
\institution{IBM Research}
\country{USA}
}
\author{Casey Dugan}
\email{cadugan@us.ibm.com}
\affiliation{
\institution{IBM Research}
\country{USA}
}
\renewcommand{\shortauthors}{Wang and Wang et al.}

\begin{abstract}
Computational notebooks allow data scientists to express their ideas through a combination of code and documentation. 
However, data scientists often pay attention only to the code, and neglect creating or updating their documentation during quick iterations. 
Inspired by human documentation practices learned from 80 highly-voted Kaggle notebooks, we design and implement Themisto, an automated documentation generation system to explore \revision{how human-centered AI systems can support human data scientists} in the machine learning code documentation scenario. 
Themisto facilitates the creation of documentation via three approaches: a deep-learning-based approach to generate documentation for source code, a query-based approach to retrieve online API documentation for source code, and a user prompt approach to nudge users to write documentation. 
We evaluated Themisto in a within-subjects experiment with 24 data science practitioners, and found that automated documentation generation techniques reduced the time for writing documentation, reminded participants to document code they would have ignored, and improved participants' satisfaction with their computational notebook.

\end{abstract}




\maketitle
\section{introduction}
Documenting the story behind code and results is critical for data scientists to collabrate effectively with others, as well as \revision{their future selves~\cite{kery_exploring_2017, piorkowski2021ai, kross2021orienting,Muller:2019:DSW:3290605.3300356}}. 
The story, code, and computational results together construct a computational narrative.
Unfortunately, data scientists often write messy and drafty analysis code in computational notebooks as they need to quickly test hypotheses and experiment with alternatives.
It is a tedious process for data scientists to then manually document and refactor the raw notebook into a more readable computational narrative, thus many people neglect to do so~\cite{rule2018exploration}.

Many efforts have sought to address the tension between \emph{exploration} and \emph{explanation} in computational notebooks.
For example, researchers have explored the use of code gathering techniques to help data scientists organize cluttered and inconsistent notebooks~\cite{Head2019CHI}, as well as algorithmic and visualization approaches to help data scientists forage past analysis choices~\cite{Kery2019CHI}.
But these efforts focus on the cleaning and organizing of existing notebook content, instead of creating the new content. 
Another work developed a chat feature that enables data scientists to have simultaneous discussions while coding in a notebook \cite{wang2020callisto}, and linked their chat messages as documentations to relevant notebook elements as in Google Docs~\cite{wang2016people}. 
However, these chat messages are too fragmented and colloquial to be used for documentation; besides, in real practice \revision{data scientists and business analysts rarely work on notebooks at the same time and actively message each other}. 

We began our project by asking, 
``What makes a well-documented notebook?'' 
To answer this question, we first conducted an in-depth analysis of how human data scientists document notebooks. 
Publicly shared user notebooks on Githubs are often not well documented~\cite{rule2018exploration}, thus we look up to a special set of notebooks -- the highly-voted notebooks users submitted to Kaggle competitions.
We conducted a formative study with a sample of 80 of these notebooks, and our interative indepth coding analysis suggested these 80 notebooks have much better documentations in comparison to the corpus reported in previous literature~\cite{rule2018exploration}. 
Thus, we refer to them as ``well-documented'' notebooks.

Our coding process of these 80 notebooks also revealed a taxtonomy of nine categories (e.g., Reason, Process, Result) for the documentation content, 
which reflects the thought processes and decisions made by the notebook owner. 
These findings together with the insights from related work motivate us to consider AI automation as a potential solution to support the human process of crafting documentation. 

We propose \sys{}, an automated code documentation generation system that integrates into the Jupyter Notebook environment.
To support the diverse types of documentation content and to complement the AI limitations, \sys{} incorporate three distinct approaches: a deep-learning-based approach to automatically generate new documentation for source code (fully automated); a query-based approach to retrieve existing documentation from online \ac{API} websites for third party packages and libraries (fully automated); and a prompt-based approach to give users a start of the sentence and encourage them to complete the sentence that serves as documentation (semi automated).

We evaluated \sys{} in a within-subjects experiment with 24 data science practitioners. 
We found that \sys{} reduced the time for data scientists to create documentation, reminded them to document code they would have ignored, and improved their satisfaction with their computational notebooks. 
Meanwhile, the quality of the documentation produced with \sys{} are about the same as what data scientists produced on their own.
\revision{Base on these findings, we re-imagine that the code documentation task can be conducted in a Human-AI Collaboration fasion in the future, where this joint effort may have unique advantages in comparison to the solo effort of a human alone.}

Our paper provides a three-fold contribution to the HCI and data science practitioner communities:
\begin{itemize}
    \item providing an empirical understanding of best practices of how human documenting a notebook through an analysis of highly-rated Kaggle notebooks,
    \item demonstrating the design of \revision{a human-centered} AI system that can collaborate with human data scientists to create high-quality computational narratives, 
    \item reporting empirical evidence that \sys{} can collaborate with data scientists to generate high quality and highly-satisfied computational notebooks in much less time.
\end{itemize}

\section{Related Work}
Our work builds on top of both \ac{HCI} and \ac{ML} fields. Thus, our literature review briefly summarizes the work of both, with a focus on the following three topics: computational notebook management, code documentation supporting systems, and neural-network-based code summarization.
\begin{figure}[t]
    \includegraphics[width=\textwidth]{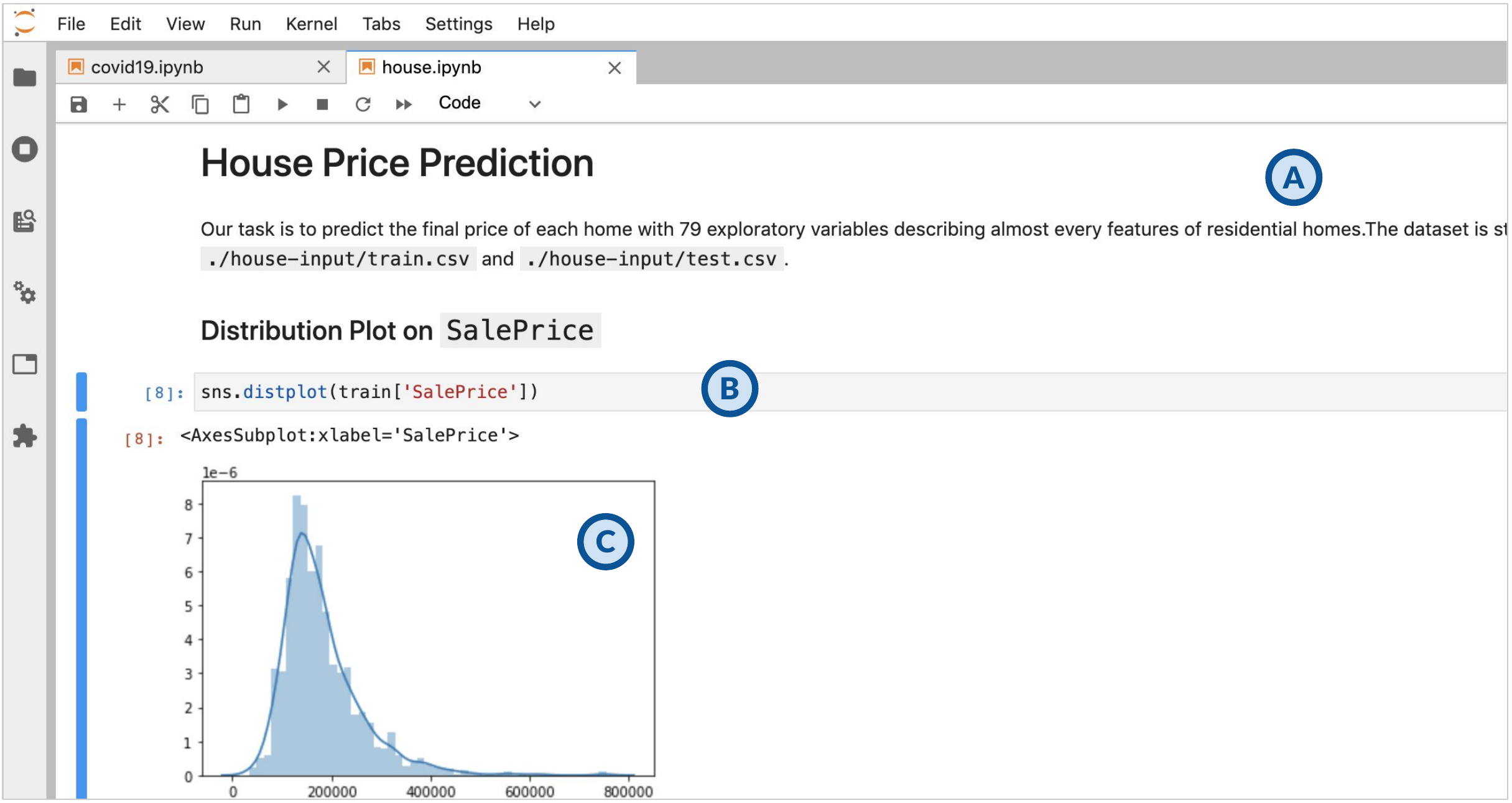}
    \caption{Computational notebooks allow data scientists to create (A) markdown cells and (B) code cells, and view (C) code output in the same environment.
    Together, the variety of media including text explanations, graphs, forms, interactive visualizations, code segments and their outputs --- weaves into computational narratives.}
    \label{fig:notebook}
\end{figure}
\subsection{Computational Notebooks as Computational Narrative}
Computational notebooks allow data scientists to weave together a variety of media, including text explanations, graphs, forms, interactive visualizations, code segments, and their outputs, into computational narratives (as shown in Figure \ref{fig:notebook}). 
These computational narratives enable literate programming~\cite{knuth_literate} and allow data scientists to effectively create, communicate, and collaborate on their analysis work. 
The data scientist community has widely adopted notebook systems (e.g., Jupyter Notebook~\cite{jupyter_project_2015} and Jupyter Lab~\cite{jupyter_jupyterlab:_2016}) as their main working envrioment~\cite{perkel_why_2018}.

However, it is not easy for data scientists to create a computational narrative while they are coding for rapid exploration.
Data scientists often need to explore diverse sets of hypotheses and theories~\cite{liu2019understanding, rehman2019towards}.
Active exploration of alternatives increases the workload for data scientists to track the history of their experimentation~\cite{Kery:2018:SNE:3173574.3173748}. 
Thus, documenting for those alternatives will pose more workload to data scientists, sometimes interference with their cognitive process of coding, and is hardly rewarding because many those alternatives will be discarded in later versions.

Because creating and maintaining a clean computational narrative is often an expensive and tedious process, many computational notebooks shared within open communities are not appropriately documented. 
For example, Rule et al. examined 1 million open-source computational notebooks from Github and found that one in four lacked any sort of written documentation~\cite{rule2018exploration}. 
In addition, they analyzed a sample of 221 academic computational notebooks, which they considered are higher quality notebooks, and found that academic computational notebooks contained text cells for introduction, describing analytical steps, explaining the reasoning, and discussing results.

Poor documentation hinders the readability and reusability of the notebooks that are shared with other collaborators or even with one's future self~\cite{chattopadhyay2020s}.
Recently, various groups of researchers have developed a wide range of tools to help data scientists to manage their ``messy'' computational notebooks.
\revision{Notably, Lau et al. summarized the design space of computational notebooks which covered an overview for improving explanations in computational notebooks~\cite{lau2020design}.
Many strategies interact with markdown comments.
For example, facilitating cell folding could help surface important markdown cells \cite{rule_aiding_2018}; 
Kery et al. designed Verdant~\cite{Kery2019CHI}, a lightweight local versioning plugin for Jupyter Lab, that uses algorithmic and visualization techniques for data science workers to better forage their past analysis choices; 
Woods et al argued \cite{wood2018design} for simpler and richer narratives;
Head et al. used code gathering tools to help data scientists trace back to the computational code from an end result~\cite{Head2019CHI};
Wenskovitch et al. designed an interactive tool that produced a visual summary of the structure of a computational notebook~\cite{wenskovitch2019albireo};
Wang et al. proposed capturing the contextual connections between notebook content and discussion messages to help data science teams reflect on their decision making process~\cite{wang2020callisto}.
}

However, despite the wide variety of approaches to helping data scientsts manage their notebooks, none of these tools directly aids data scientists in creating new, rich, descriptive contents to document their computational notebooks, and to improve the quality of the computational narrative. 
Recent research works have proposed to use AI solutions to automate the various tasks along a data science project, such as the model training, model selection, and feature selection, and these technology are commonly refered as AutoML~\cite{wang2021autods,liu2020admm}.
The research gap and the AutoML techniques motivate us to design and build an AI system to support data scientists to better document their code and to produce higher qualitive computational narratives.

But what makes up a good computational narrative? 
Despite the portrait of not-so-good notebooks on Github \cite{rule2018exploration}, we need further understanding and role models for well-documented computational narratives. 
Thus, we decided to first conduct an in-depth analysis of some highly-voted notebooks on Kaggle competetion\footnote{https://www.kaggle.com/}.
Kaggle competition provides a platform where organizations post datasets as challenges, and many data scientists submit their notebooks as solutions to a challenge.
If a solution has the highest accuracy, it wins the competition. 
But those winning solutions are often not the most voted ones, as community members voted on readability and completeness of the computational narrative.

\subsection{Code Documentation in Software Engineering}
Documentation plays an important role in software programming. 
Programmers write comments in their source code to make the code easier for both themselves and others to understand \cite{padioleau2009listening}.
Writing clear and comprehensive documentation is critical to software development and maintenance~\cite{de2005study, kajko2005survey, shi2011empirical, roehm2012professional, maalej2013patterns}. 
However, writing documentation itself is a time-consuming task. And that is why documentation practices in open source communities are widely perceived to be of low quality, due in part to low levels of intrinsic enjoyment for doing documentation work~\cite{geiger2018types}.

To save time in creating documentation, template-based approaches are often used to help developers annotate their source code.
For example, tools like JavaDoc\footnote{JavaDoc: \url{https://docs.oracle.com/javase/8/docs/technotes/tools/windows/javadoc.html}} and JSDoc\footnote{JSDoc: \url{https://devdocs.io/jsdoc/}} allow programmers to annotate their code with tags (e.g., @param, @return) and then automatically generate documentation using these tags. 
This approach helps programmers create documentation for others and works especially well for documenting \acp{API}, where method signatures and variable types are important pieces of information and can easily be documented from tags.
Although, such methods may not work in the rapid, experimental nature of data science work, because data scientists may be particularly reluctant to create and maintain high-quality documentation of their work. 
Furthermore, these methods can not capture other aspects of documentation important in data science, such as how a data set was constructed, the intent behind an analysis, or a description of why an experiment was successful or not.

Recently, some researchers have put forth proposals for better documenting the specific artifacts invovled in a data science workflow, i.e., the data set and the machine learning model~\cite{gebru2018datasheets, holland2018dataset, mitchell2019model, arnold2019factsheets}.
Notably, Gebru et al.~\cite{gebru2018datasheets} and Holland et al.~\cite{holland2018dataset} proposed both qualitative and quantitative \revision{guidelines} for documenting a dataset, so that the dataset creators and maintainers can follow these guidlines to document useful information for the data. 
Similarly, Mitchell et al.~\cite{mitchell2019model} and Arnold et al.~\cite{arnold2019factsheets} explored using formulas to document the machine learning model artifacts, and sharing such formulas with others.
These approaches are inline with what is called \textit{provenance}, which refers to tracking what has been done with code and data over time -- typically to aid reproducibility of results -- using applications such as noWorkflow and YesWorkflow \cite{pimentel2016yin}, ReproduceMeGit \cite{samuel2020reproducemegit}, and Provbook \cite{samuel2018provbook}.
However, this approach focus more on the the dataset and the model artifact in the final product of a data science project, and supporting data scientists to create a ``factsheet'' for these artifacts for the non-technical consumers. Instead, we want to support data scientists to better create the documentation during the process of creating models and data science products, and such documentation, together with the code as a computational narrative, is primarily for other technical users to understand and to reuse.

In addition to the various ways of generating new documentation for code, there is another research line that focuses on improving the usability of documentation, as novice programmers may find it difficult to read and use \ac{API} documentation~\cite{horvath2019methods}.
Oney et al. proposed linking interactive documentation and example code in an editor to help novice programmers better understand the external documentation and write code~\cite{oney2012codelets}.
We believe this approach of linking code with external documentation is a promising way to help data scientists to create more usable documentations, and we will also implement this retrieval-based approach in our system.

\subsection{GNN-Based Automatic Code Summarization System}
Automatic code summarization is a rapidly-expanding research topic in the \ac{NLP} and \ac{ML} communities~\cite{eddy2013evaluating, hu2018deep, alon2018code2seq, iyer_summarizing_2016, leclair2020improved}.
The automatic code summarization task can be considered as a translation task, which takes a code snippet as the input sequence, and \say{translating} it into a natural language description of the code as an output sequence.

Early work primarily used predefined templates and heuristics to produce code summaries (e.g.,~\cite{eddy2013evaluating, sridhara2010towards}). 
Recent studies have taken the advantage of modern deep neural network architectures to generate the summary for the code (e.g.~\cite{hu2018deep, alon2018code2seq, iyer2016summarizing}).
Motivated by the language translation task (e.g., English to French), most of these learning-based approaches are based on the \ac{NMT} model architecture~\cite{luong2015effective}. 
This architecture breaks code into a sequence of input tokens and produce the summarization text as a sequence of output tokens. 

However, this sequence-to-sequence approach does not work well in practice because source code is not just a stream of tokens.
There is additional semantic information that is lost when processing source code in this way. 
LeClair et al. proposed improving code summarization through the use of \acp{GNN}~\cite{leclair2020improved}.
The GNN model can take in both the code sequence and its \ac{AST} structure (refer to Fig~\ref{fig:model} for an example) as input to generate summary sentences as output.
Their approach achieved better accuracy than the baseline algorithms.

Our work explores neural-network-based automatic code summarization techniques to support document writing in computational notebooks.  
To our acknowledge, there has been little discussion in the \ac{HCI} community on leveraging automatic code summarization techniques to improve documentation.
Furthermore, we suspect that the automation approach alone may not work well in the documentation creation task, as data science is a highly interdisciplinary field that requires various human expertises to explain and interpret. 
Inspired by prior studies that implement \ac{AI} systems to work together with human~\cite{louie2020novice, wang2020autoai}, we believe the system will work better if it has both the automated documentaiton capability and the capability that allows users to directly manipulate the documentation.
However, what types of documentation may be better suited for \ac{AI} to do, and what works should the system leave to human data scientists? 
This is a design question that requires further exploration of the best practices for creating notebooks. 
Thus, we start this project with a formative study to fill this research gap.
\section{Formative Study}
In order to build a useful system that can support data scientists to create documentations and to improve their computational narrative's quality, we first need to explore and understand the characteristics of good documentations in high-quality notebooks. \textbf{What does a well-documented computational narrative look like?}   
We identify \say{well-documented} computational narratives with ratings from a broader data scientist community (Kaggle), and analyze their characteristics specifically around the documentation.
We consider the community voting number is a good indicator to reflect a computational notebook's quality for our research goal. 
Based on this premise, we then conduct a formative study to analyze the characteristics of a set of most voted computational narratives, and explore how the data scientists create documentations for these notebooks.

\subsection{Data Collection}
We collected notebooks from two popular Kaggle competitions --- House Price Prediction\footnote{\url{https://www.kaggle.com/c/house-prices-advanced-regression-techniques}} and Titanic Survival Prediction\footnote{\url{https://www.kaggle.com/c/titanic/}}.
We chose these two competitions because they are the most popular competitions (5280 notebooks submitted for House Price and 6300 notebooks submitted for Titanic Survival) and because many data science courses use these two competitions as a tutorial for beginners~\cite{bai2018problem, fernandez7energy}.

We collected the top 1\% of the submitted notebooks from each competition based on their voting numbers, which resulted in 53 for House Price and 63 for Titanic Survival.
We then filtered out the notebooks that were not written in English and the ones that are not relevant to the particular challenge (e.g., a computational notebook as a tutorial on how to save memories can win lots of votes, but it is not a solution to the challenge), which returned 80 valid notebooks for analysis (39 for House Price and 41 for Titanic Survival).
\begin{figure}[t]
    \includegraphics[width=\textwidth]{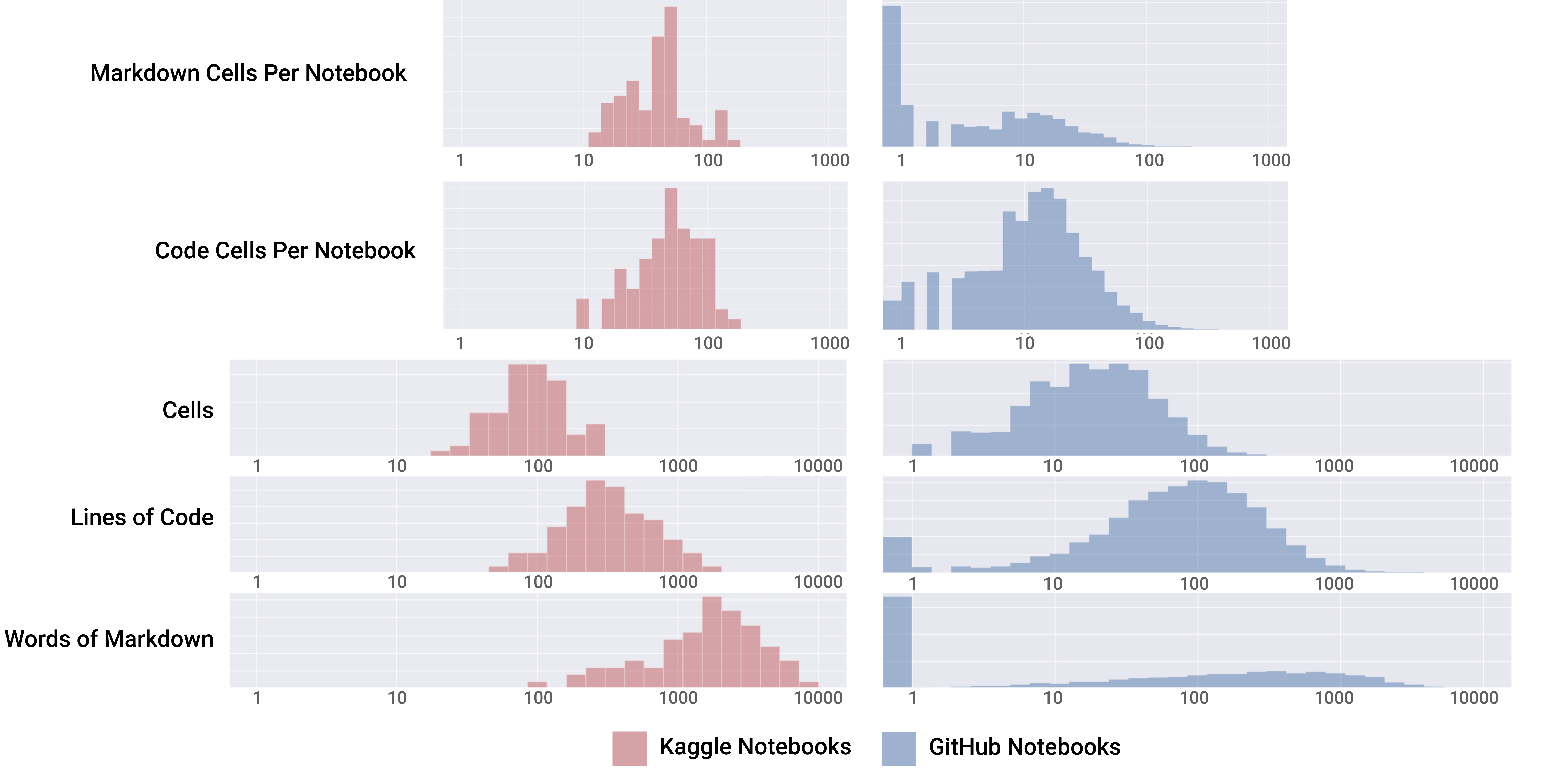}
    \caption{\revision{We replicated the notebook-level descriptive analysis by Rule et al. \cite{rule2018exploration} to the 80 well-documented notebooks on Kaggle. The left side represents the descriptive visualization of the 80 well-documented computational notebooks from Kaggle (noted as Sample A) and the right side represents the descriptive visualization of the 1 million computational notebooks on Github (noted as Sample B). The highly-voted notebooks on Kaggle are better documented compared to the Github notebooks.}}
    \label{fig:rule}
\end{figure}

\subsection{Data Analysis}
Five members of the research team conducted an iterative open coding \revision{process} to analyze the collected notebooks. 
Differing from ~\cite{rule2018exploration}, where their qualitative coding \revision{stopped} at the notebook level, our analysis goes deep to the cell granularity: we code each cell's purposes and types of content; and which step (stages) in the data science lifecycle that the cell belongs (e.g., data cleaning or modeling training~\cite{zhang2020data,automationsurvey}).
Our analysis covered 4427 code cells and 3606 markdown cells within the 80 notebooks.
Each notebook took around 1 hour to code as we coded the notebook at the cell level.

Each coder independently analyzed the same six notebooks to develop a codebook. 
After discussing and refining the codebook, they again went back to recode the six notebooks and achieved pair-wise inter-rater reliability ranged 0.78-0.95 (Cohen's $\kappa$).
After this step, the five coders divided and coded the remaining notebooks.

\begin{table}[t!]
    \caption{We identified 9 categories based on the purpose of markdown cells. Note that a markdown cell may belong to multiple categories of contents or none of the categories.}
    \label{tab:category}
    \small
    \begin{tabular}{p{1.8cm} p{1cm} p{5cm} p{4cm}}
        \toprule
        \textbf{Category} & \textbf{N} & \textbf{Description} & \textbf{Example} \\
        \midrule
        Process & 2115 \newline (58.65\%) & The markdown cell describes what the following code cell is doing. This always appears before the relevant code cell. & \mintinline{html}{Transforming Feature X}\newline \mintinline{html}{to a new binary variable} \\
        \midrule
        Headline & 1167 \newline (32.36\%) & The markdown cell contains a headline in markdown syntax. The cell is used for navigation purposes or marking the structure of the notebook. It may be relevant to a nearby code cell. & \mintinline{html}{# Blending Models}\\
        \midrule
        Result & 692 \newline (19.19\%) & The markdown cell explains the output. This type always appears after the relevant code cell. & \mintinline{html}{It turns out there is a}\newline \mintinline{html}{long tail of outlying}\newline \mintinline{html}{properties...} \\
        \midrule
        \revision{Background Knowledge} & 414 \newline (11.48\%) & The markdown cell provides a rich content \revision{for background knowledge}, but may not be relevant to a specific code cell. & \mintinline{html}{Multicollinearity incre-} \newline \mintinline{html}{ases the standard errors} \newline \mintinline{html}{of the coefficients.} \\
        \midrule
        Reason & 227 \newline (6.30\%) & The markdown cell explains the reasons why certain functions are used or why a task is performed. This may appear before or after the relevant code cell. & \mintinline{html}{We do this manually, be-}\newline\mintinline{html}{cause ML models won't be}\newline\mintinline{html}{able to reliably tell}\newline\mintinline{html}{the differences.} \\
        \midrule
        Todo & 202 \newline (5.60\%) & The markdown cell describes \revision{a list of actions for upcoming analysis. This normally is not relevant to a specific code cell.} & \mintinline{html}{1. Apply models}\newline\mintinline{html}{2. Get cross validation}\newline\mintinline{html}{scores} \newline \mintinline{html}{3. Calculate the mean} \\
        \midrule
        Reference & 200 \newline (5.55\%) & The markdown cell contains an external reference. This is also relevant to the adjacent code cell. & \mintinline{html}{Gradient Boosting} \newline\mintinline{html}{Regression Refer}\newline\mintinline{html}{[here](https://...)}\\
        \midrule
        Meta-Information & 141 \newline (3.91\%) & The markdown cell contains meta-information such as project overview, author's information, and a link to the data sources. This often is not relevant to a specific code. & \mintinline{html}{The purpose of this}\newline \mintinline{html}{notebook is to build a}\newline \mintinline{html}{model with Tensorflow.}\\
        \midrule
        Summary & 51 \newline (1.41\%) & The markdown cell summarizes what has been done so far for a section or a series of steps. This often is not relevant to a specific code.&\mintinline{html}{**In summary**}\newline\mintinline{html}{By EDA we found a strong}\newline\mintinline{html}{impact of features like}\newline\mintinline{html}{Age, Embarked..} \\
        \bottomrule
    \end{tabular}
\end{table}

\subsection{Results}
We found that these 80 well-documented computational notebooks all contain rich documentation.
In total, we identified \textbf{nine categories for the content} of the markdown cells.
In addition, we found the markdown cells covered \textbf{four stages and 13 tasks} of the data science workflow~\cite{automationsurvey}.
Note that a markdown cell may belong to multiple categories.

\subsubsection{Descriptive statistics of the notebook.}
We found that on average, each notebook contains 55.3 code cells and 45.1 markdown cells.
We replicated \revision{the} notebook descriptive analysis that Rule et al. used to analyze 1 milion computational notebooks on Github \cite{rule2018exploration}.
As shown in Figure \ref{fig:rule}, the left side represents the descriptive visualization of the 80 well-documented computational notebooks from Kaggle (noted as Sample A) and the right side reprersents the descriptive visualization of the 1 milion computational notebooks on Github (noted as Sample B).
We found that the Sample A has more total cells per notebook (Median = 95) than Sample B (Median = 18).
Sample A has roughly equal ratio of markdown cells and code cells per notebook, while Sample B is unbalanced with majority cells being code cells.
Notably, Sample A has more total words in markdown cells (Median = 1728) than Sample B (Median = 218). 
This result indicates that the 80 well-documented computational notebooks are better documented than general Github notebooks.

\subsubsection{Data scientists use markdown cells to document a broad range of topics.}
As shown in Table \ref{tab:category}, our analysis revealed that markdown cells are mostly used to describe what the adjacent code cell is doing (Process, 58.65\%).
Second to the Process category, 32.36\% markdown cells are used to specify a headline for organizing the notebook into separate functional sections and for navigation purposes (Headline).

Markdown cells can also be used to explain beyond the adjacent code cells. 
We found that many markdown cells are created to describe the outputs from code execution (Result, 19.19\%), to explain results or critical decisions (Reason, 6.30\%), or to provide an outline for the readers to know what they are going to do in a list of todo actions (Todo, 5.60\%), and/or to recap what has been done so far (Summary, 1.41\%).

We observed that 11.48\% markdown cells explain what a general data science concept means, or how a function works (\revision{Background Knowledge}), while 5.54\% markdown cells are connected with external references for readers to further explore the topics (Reference). 
We believe these are the extra efforts that the notebook owners dedicated, to attract a broader audience, especially beginners in the Kaggle community.
In addition, we found that authors approached the story in different styles.
For example, some authors want to leave their own signature, and so they spend spaces at the beginning of the notebooks to debrief the project, to add the author's information, or even to add their mottos (Meta-Information, 3.91\%).
Some authors prefer to use concise and accurate language to convey important information; while others write documentation in more creative and entertaining ways --- for example, making analogies between data science workflow and starting a business.


\begin{table}[t!]
\caption{We coded each markdown cell to which data science stage (or task) they belong. We identified 4 stages with 13 tasks out of the data science lifecycle~\cite{automationsurvey}. Note that a markdown cell may belong to multiple stages or none of the stages.}
\begin{center}
\small
\begin{tabular}{llp{4.5cm}l}
	\toprule
	\textbf{Stage} & \textbf{Total} & \textbf{Task} & \textbf{N} \\
	\midrule
	\multirow{2}{*}{Environment Configuration} &	\multirow{2}{*}{162 (4.49\%)} & Library Loading & 33 (0.92\%)\\
	&&Data Loading & 129 (3.58\%)\\
	\midrule
	\multirow{3}{*}{Data Preparation and Exploration}&
	\multirow{3}{*}{1336 (37.05\%)}&
	Data Preparation & 91 (2.52\%)\\
	&&Exploratory Data Analysis & 960 (26.62\%)\\
	&&Data Cleaning & 285 (7.90\%)\\
	\midrule
	\multirow{3}{*}{Feature Engineering and Selection}& 
	\multirow{3}{*}{375 (10.40\%)}&
	Feature Engineering & 120 (3.32\%)\\
	&& Feature Transformation & 178 (4.94\%)\\
	&&Feature Selection & 77 (2.14\%)\\
	\midrule
	\multirow{5}{*}{Model Building and Selection}&
	\multirow{5}{*}{994 (27.57\%)}&
	Model Building & 247 (6.85\%)\\
	&& Data Sub-Sampling and Train-Test Splitting & 61 (1.69\%)\\
	&& Model Training & 377 (10.45\%)\\
	&& Model Parameter Tuning & 81 (2.25\%)\\
	&& Model Validation and Assembling & 288 (6.32\%)\\
	\bottomrule
\end{tabular}
\end{center}
\label{tab:stage}
\end{table}

\subsubsection{Data science stages.}
We coded markdown cells based on where they belong in the data science workflow \cite{wang2019human}.
\revision{As shown in Table \ref{tab:stage},} we identified four stages and 13 tasks. 
The four stages include \textbf{environment configuration} (4.50\%), \textbf{data preparation} and exploration (37.05\%), \textbf{feature engineering and selection} (10.40\%), and \textbf{model building and selection} (27.57\%).
At the finer-grained task level, in particular, notebook authors create more markdown cells for documenting exploratory data analysis tasks (26.62\%) and model training tasks (10.45\%).
The rest of the markdown cells are evenly distributed along with other tasks.

\subsection{Design Implications}
In summary, our analysis of markdown cells in well-documented notebooks suggests that data scientists document various types of content in a notebook, and the distribution of these markdown cells generally follows an order of the data science lifecycle, starting with data cleaning, and ending with model building and selection. Based on these findings, we synthesize the following actionable design considerations:

\begin{itemize}
    \item \textbf{The system should support more than one type of documentation generation.} Data scientists benefit from documenting not only the behavior of the code, but also interpreting the output, and explaining rationales. Thus, a good system should be flexible to support more than one type of documentation generation.
    \item \textbf{Some types of documentations are highly related to the adjacent code cell.} We found at least the Process, Result, Reason, and Reference types of documentations are highly related to the adjacent code cell. To automatically generate interpretations of results or rationale for a decision may be hard, as both involve deep human expertise. But, with the latest neural network algorithms, we believe we can build an automation system to generate Process type of documentation, and we can also retrieve Reference for a given code cell.
    \item \textbf{There are certain types of documentations that are irrelevant to the code.} Various types of documentations do not have a relevant code piece upon which the automation algorithm can be trained. Together with the Reason and Result types, the system should also provide a function that the human user can easily switch to the manual creation mode for these types.
    \item \textbf{For different types of documentation, it could be at the top or the bottom of the related code cell.} This design insight is particularly important to the Process, Result, and Reason types of documentation. It may be less preferable to put Result documentation before the code cell, where the result is yet to be rendered. The system should be flexible to render documentation at different relative locations to the code cell.
    \item \textbf{External resources such as \acp{URL} and the official \ac{API} descriptions may also be useful.} Some types of documentation, such as \revision{Background Knowledge} and Reference, are not easy to be generated with the NN-based models, but they are easy to retrieve from the Internet. So the system should incorporate the capability to fetch relevant web content as candidate documentation.
    \item \textbf{There is an ordinality in markdown cells that is aligned with the data science project's lifecycle.} The system should consider that Library Loading types of cells are often at the beginning section of the notebook, and the Model Training type of content may be more likely to appear near the end of the notebook. In our system prototype, though, we did not take this design consideration into account, it will be our future work.
    \item \textbf{The notebook would be nice to have documentation with a problem overview at the beginning and a summary at the end.} We considered this design implication not in the system design, but our evaluation study design. For the two barebone notebooks we used in the experiment, we always provide a problem overview as a markdown cell at the top of the notebook.
\end{itemize}
\section{The Themisto System: Design and Implementation}
Based on findings from the formative study and design insights from related works, we design and implement \sys{}, an automatic documentation generation system that supports data scientists to write better-documented computational narratives. 
In this section, we present the system architecture, the user interface design, and the core technical capability of generating documentation. 

\subsection{System Architecture}
The \sys{} system has two components: the client-side \ac{UI} is implemented as a Jupyter Notebook plugin using TypeScript code, and the server-side backend is implemented as a server using Python and Flask.

The client-side program is responsible to render user interface, and also to monitor the user actions on the notebook to edits in code cells.
When the user's cursor is focused on a code cell, the \ac{UI} will send the current code cell content to the server-side program through \ac{HTTP} requests.

The server-side program takes the code content and generates documentation using both the deep-learning-based approach and the query-based approach.
For the deep-learning-based approach, the server-side program first tokenizes the code content and generates the \ac{AST}.
It then generates the prediction with the pre-trained model.
For the query-based approach, the server-side program matches the curated \ac{API} calls with the code snippets and returns the pre-collected descriptions.
For the prompt-based approach, the server-side program sends different prompts (e.g., for interpreting result or for explaining reason) base on the output type of the code cell.


\subsection{User Interface Design}

\begin{figure}[t]
    \includegraphics[scale=0.5]{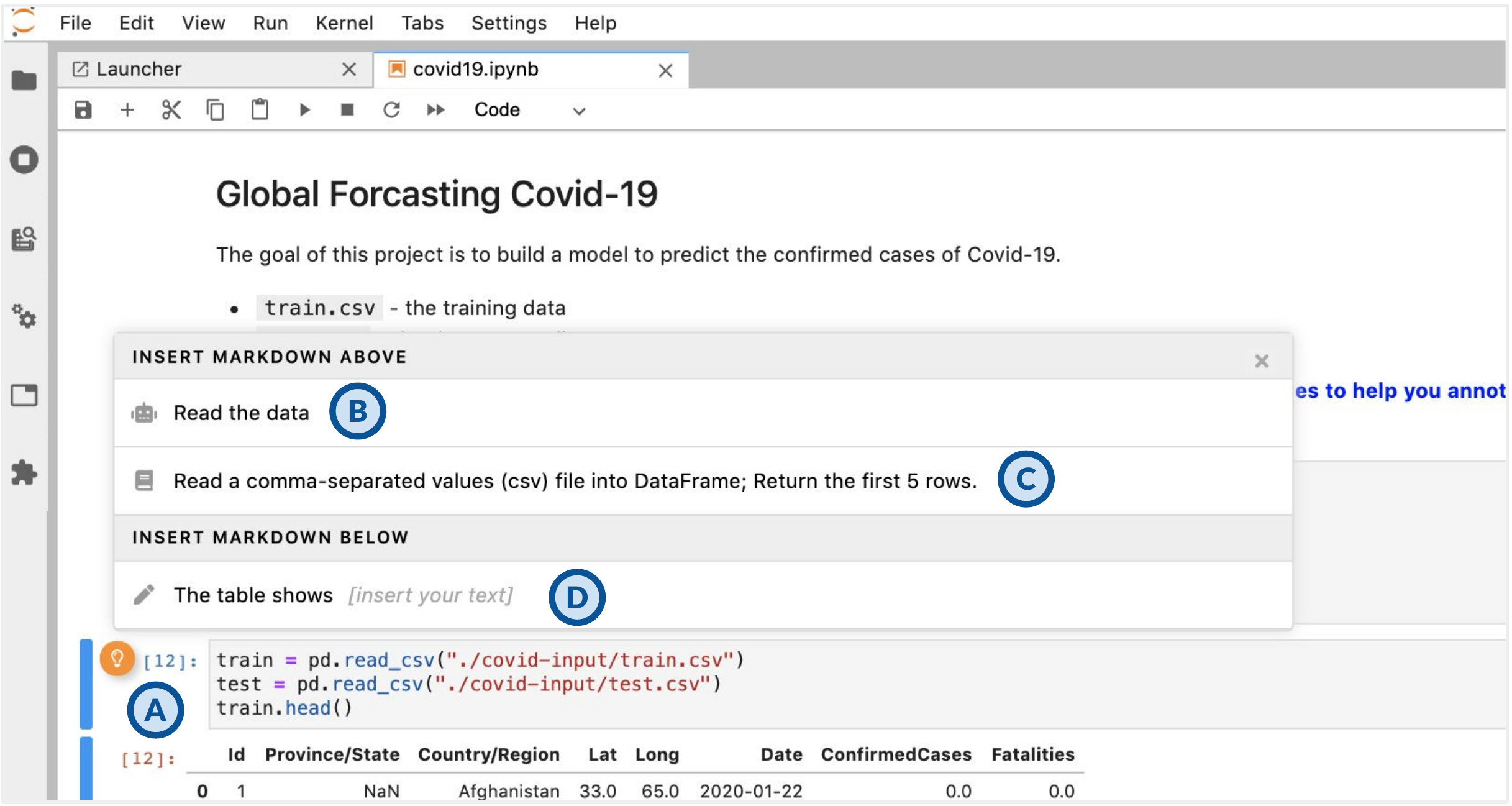}
    \caption{The \sys{} user interface is implemented as a Jupyter Notebook plugin: (A) When the recommended documentation is ready, a lightbulb icon shows up to the left of the currently-focused code cell. (B -- D) shows the three options in the dropdown menu generated by \sys{},  (B) A documentation candidate generated for the code with a deep-learning model, (C) A documentation candidate retrieved from the online \ac{API} documentation for the source code, and (D) A prompt message that nudges users to write documentation on a given topic.}
    \label{fig:interface}
\end{figure}

Figure \ref{fig:interface} shows the user interface of \sys{} as a Jupyter Notebook plugin. 
Each time the user changes their focus on a code cell, as they may be inspecting or working on the cell, the plugin is triggered. 
The plugin sends the user-focused code cell's content to the backend. Using this content, the backend generates a code summarization using the model and retrieves a piece of documentation from the \ac{API} webpage. 
When such a documentation generation process is done, the generated documentation is sent from the server-side to the frontend, and a light bulb icon appears next to the code cell, indicating that the there are recommended markdown cells for the selected code cell (as shown in Figure \ref{fig:interface}.A).

When a user clicks on the light bulb icon \revision{which appears next to any selected code cells, \sys{} render all the three options in the dropdown menu}: (1) a deep-learning-based approach to generate documentation for source code (Figure \ref{fig:interface}.B); (2) a query-based approach to retrieve the online \ac{API} documentation for source code (Figure \ref{fig:interface}.C); and (3) a user prompt approach to nudge users to write more documentation (Figure \ref{fig:interface}.D).
If the user likes one of these three candidates, they can simply click on one of them, and the selected documentation candidate will be inserted into above the code cell (if it is the Process, Reference, or Reason type), or below it (if it is the Result type).

\begin{figure}[t]
    \includegraphics[width=\columnwidth]{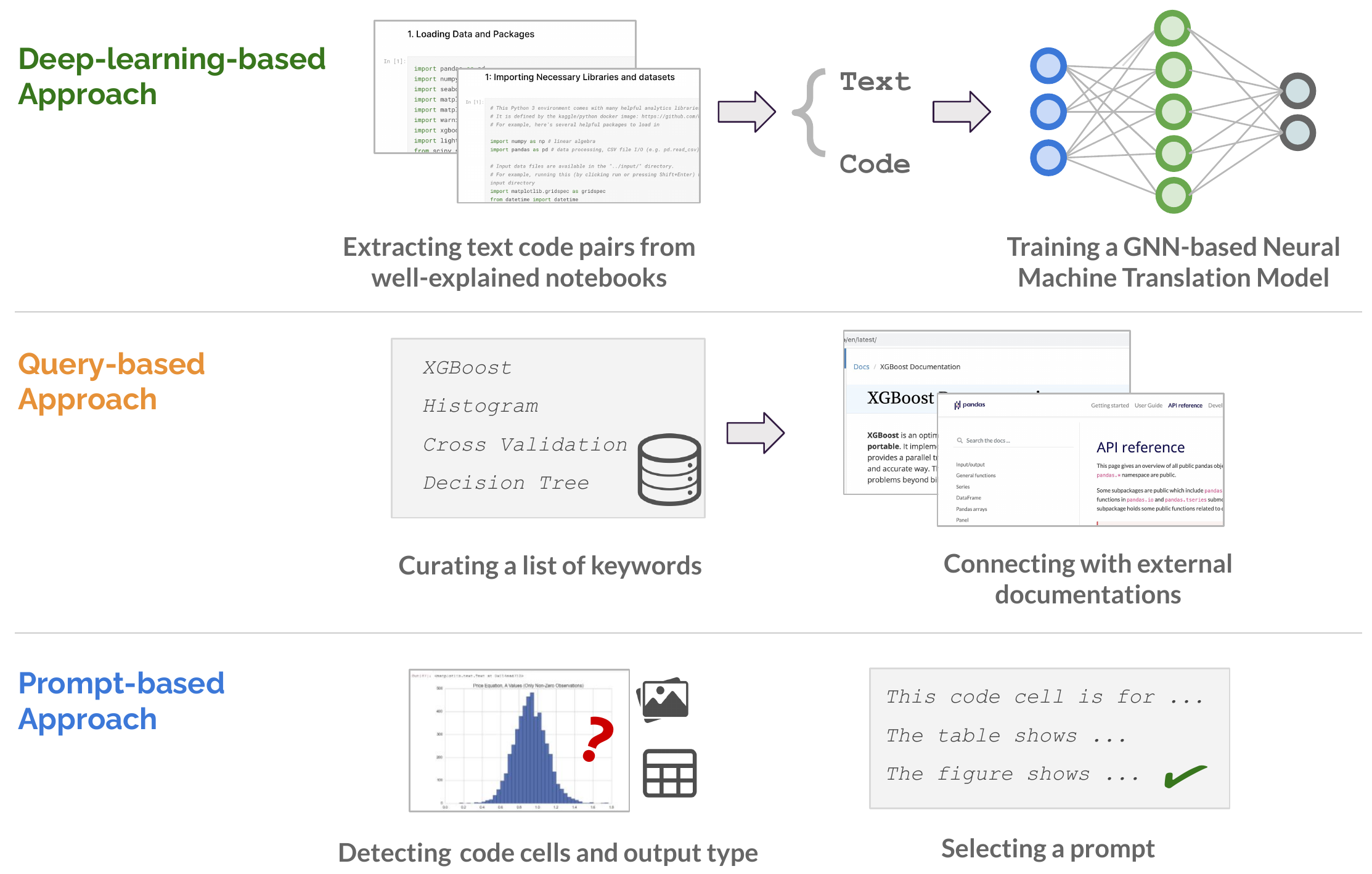}
    \caption{An illustration of the three different approaches for documentation generation in \sys{}.}
    \label{fig:workflow}
\end{figure}

\subsection{Three Approaches for Documentation Generation}
In this subsection, we describe the rationale and implementation detail of the three different approaches for documentation generation (Figure \ref{fig:workflow}):
\begin{itemize}
    \item Our formative study suggests that the system should be able to \textbf{generate multiple types of documentation} (e.g., Process, Result, \revision{Background Knowledge}, Reason, and Reference).
    \item Some types of documentation can be \textbf{directly derived from the code}, thus the automated approaches can help. The Process type of documentation directly describes the coding process, and existing ML literature suggest that the deep-learning-based approach is most suitable for generating it; The Reference type does not need a learning-based approach, it can be achieved with a traditional query-based approach, which locates and retrieves the most relevant online documentation as candidates;
    \item Some others types of documentations (e.g., Education, Result, and Reason) are \textbf{not directly related to the code}, thus the fully automated approaches are not capable of generating such contents. We design the prompt-based approach for users to complete the generation process.
\end{itemize}

\subsubsection{Deep-Learning-Based Approach}
We trained a \revision{deep learning model}\footnote{\revision{We release a larger dataset and a refined version of the model in a separate paper \cite{liu2021haconvgnn}.}} using the Graph-Neural-Network architecture based on LeClair et al. \cite{leclair2020improved}.
These \ac{GNN} models can take both the source code's structure (extracted as \ac{AST}) and the source code's content as input. 
Thus, it outperforms the traditional sequence-to-sequence model architectures, which only takes the source code's content as an input sequence, in source code summarization tasks for Python code\footnote{All the collected data science notebooks are in Python.}. We did not consider T5, BerT or GPT-3 architectures as these models can take minutes to make one inference (i.e., generate one summary) even with a cluster of GPUs (costing thousands of dollars per hour), whereas our GNN-based model can make an inference within a second with one GPU.

\begin{figure}[t]
    \includegraphics[width=\columnwidth]{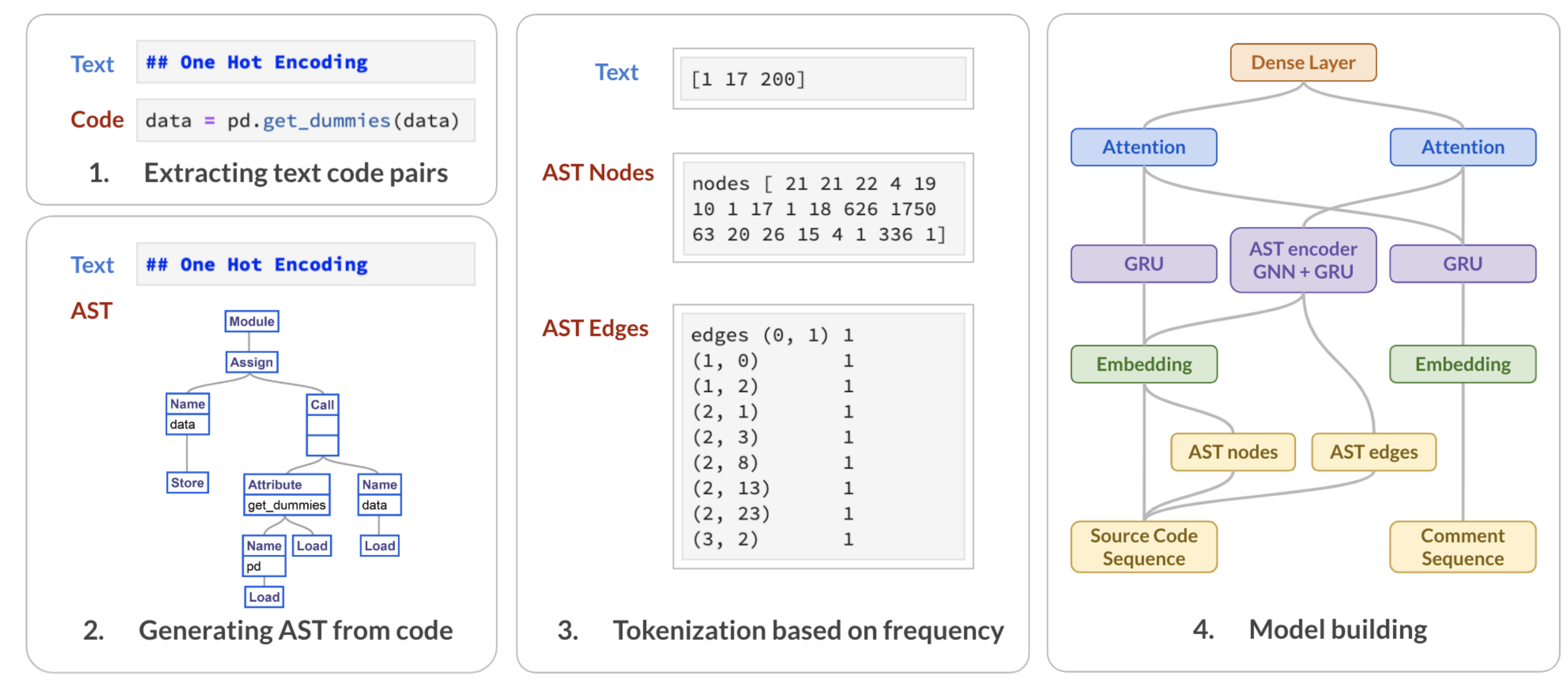}
    \caption{\revision{A code summarization model for the deep-learning-based documentation generation approach via \ac{GNN}. There are three steps of data pre-processing (1) We first extract text code pairs from existing notebooks. (2) We generate \ac{AST} from code. (3) We tokenized each word and translated them into embeddings. And (4), the \ac{GNN} model architecture.}}
    \label{fig:model}
\end{figure}

In order to fine-tune the model, 
we constructed a training dataset for our particular context. 
We collected the top 10\% highly-voted notebooks from two popular Kaggle competitions -- House Price Prediction and Titanic Survival Prediction (N = 1158). 
For each of the notebook, we first extracted code cells and the markdown cells adjacent above as a pair of input and output (similar to the data collection approach in \cite{agashe2019juice}). 
If there is an inline comment in the first line of the code cell, we replaced the output of the pair using the inline comment. 
In total, our dataset has 5,912 pairs of code and its corresponding documentation.
Following the best practice of model training, we split the dataset into training, testing, and validation subsets with an 8 to 1 to 1 ratio.

Before feeding data into the training process, we have a three-step pre-processing stage, as illustrated in Figure~\ref{fig:model}. Step 1 removes the style decoration, formats, and special characters that are not in Python grammar (e.g., Notebook Magics).
We also generate an \ac{AST} for the source code input in step 2 with Python AST library\footnote{https://docs.python.org/3/library/ast.html}. The AST result is equivalent to the source code but with more contextual and relational information.
In step 3, we tokenize source code to a sequence of tokens with an input dictionary, and parse the \ac{AST} nodes as a sequence of tokens with the same input dictionary. We parse the relationship between AST nodes as a matrix of edges. 
Finally, we tokenize the output documentation as a sequence of tokens with a separate output dictionary. 
After this process, all the tokens are transformed into an array of word embeddings --- vectors of real numbers. 
We use these data to train the network for 100 epochs, 30 batch sizes, and 15 early stop points on a two Tesla V100 GPU cluster. 
Out of all the epochs, we selected the model with the highest validation accuracy score. 

To evaluate our model's performance against baseline models, we conducted both quantitative and qualitative evaluations, as suggested by~\cite{ribeiro2020beyond}. 
For the automated quantitative evaluation, we use BLEU scores \cite{papineni2002bleu} as the model performance metric.
BLEU scores are commonly used in the source code summarization tasks. 
It evaluates the word similarity between the generated text and the ground truth text.
We selected and trained Code2Seq model~\cite{alon2018code2seq} and Graph2Seq model~\cite{xu2018graph2seq} with the same data split.

Our model achieves 11.41 (BLEU--a), which outperforms the baseline models Code2Seq (BLEU--a = 9.61) and Graph2Seq (BLEU--a = 11.05). These scores suggest that the data science documentation task is more difficult than the benchmark code summarization tasks in the software engineering field. 
For example, in data science, a notebook code cell can contain multiple code snippets and functions. 

\begin{table}[t!]
\small
\centering
\caption{\revision{Example output from the model. (Example A) The generated text well describes the code. (Example B) The generated text vaguely describes the code. (Example C) The generated text is poorly readable, but still captures the keywords of the descriptions.}}
\label{tab:example_output}
\begin{tabular}{p{2cm}p{7cm}p{3.8cm}} 
 \toprule
 \textbf{Example} & \textbf{Code Cell} & \textbf{Output From the Model}\\
 \midrule
 Example A &
\begin{tabminted}{python}
train = pd.read_csv('./house-input/train.csv')
test = pd.read_csv('./house-input/test.csv')
\end{tabminted}
& Read the data
\\
\hline
Example B &
\begin{tabminted}{python}
all_data = pd.get_dummies(all_data)
\end{tabminted} 
& Convert all the data\\
\hline
Example C &
\begin{tabminted}{python}
pred = Tree_model.predict(x_test)
pred = pd.DataFrame(pred)
pred.columns = ["ConfirmedCases_prediction"]
\end{tabminted}
& Predicate to use a predict function for tests \\
\bottomrule
\end{tabular}
\end{table}

In addition to the automated quantitative evaluation, we also conduct a qualitative analysis of the generated documentation pieces. 
We found that despite the word-to-word similarity score is low, the general quality of the content is reasonable and satisfying for building a prototype system. 
As an illustration, we provide three examples with both input and model generated outputs, as shown in Table \ref{tab:example_output}. 
In the Appendix, we provide full code cells and model-generated outputs for the two experimental notebooks that we used in the user study.


\subsubsection{Query-Based Approach}
Our formative study indicates that the well-documented Kaggle notebooks often have the description of frequently-used data science code functions for educational purposes. 
And sometimes data scientists directly paste in a link or a reference to the external \ac{API} documentation for a code function.
Thus, we implement a query-based approach that curates a list of \acp{API} from commonly used data science packages, and the short descriptions from external documentation sites.
In our system, we only cover Pandas\footnote{\url{https://pandas.pydata.org/docs/reference/index.html}}, Numpy\footnote{\url{https://numpy.org/doc/stable/reference/}}, and Scikit-learn\footnote{\url{https://scikit-learn.org/stable/modules/classes.html}} these three libraries as a starting point to explore this approach. 
We argue that it can be easily expanded to include other packages in the future.
We collected both the \ac{API} names and the short descriptions by building a crawling script with Python.
When users trigger this query-based approach for a code cell, \sys{} matches the \ac{API} names with the code snippets and concatenate all the corresponding descriptions.

\subsubsection{Prompt-Based Approach}
Lastly, the system also provides a prompt-based approach that allows users to manually create the documentation.
Because our formative study found that a well-documented notebook not only documents the process of the code, but also interprets the output, and explains rationales.
These types of documentation are hard to generate with automated solutions
To achieve it, we implement a prompt-based approach.
It detects whether the code cell has a cell output or not: if the cell outputs a result, \sys{} assumes that the user is more likely to add interpretation for the output result, thus the corresponding prompt will be inserted below the code cell. 
Otherwise, the system assumes the user may want to insert a reason or some educational types of documentations, thus it changes its prompt message.

\section{User Evaluation of Themisto}
To evaluate the usability of \sys{} and its effectiveness in supporting data scientists to create documention in notebooks, we conducted a within-subject controlled experiment with 24 data scientists. 
The task is to add documentation to the given notebook. And each participant is asked to finish two sessions, one with the \sys{} support and one without its support. 
The evaluation aims to understand (1) how well Themisto can facilitate documenting notebooks and (2) how data scientists perceive the three approaches that are used by \sys{} for generating documentation.

\subsection{Participants}
\revision{We recruited 24 data science professionals as our evaluation participants in a multinational IT company.} 
We used a snowball sampling approach to recruit participants, where we sent recruitment messages to friends and colleagues, various internal mailing-lists, and Slack channels. 
We then asked participants to refer their friends and colleagues. 
Our recruitment criteria are that the participant must have had experience in data science projects and they are familiar with Python and Jupyter Notebook environment.
As shown in Table \ref{tab:demographic}, participants reported a diverse job role backgrounds, including expert data scientists (N = 8), novice data scientists (N = 9), \acp{AIOP} or \ac{ML} engineers (N = 2), subject matter experts (N = 1), and application developer (N = 4).

\begin{table}[t!]
    \caption{Demographics of participants}
    \label{tab:demographic}
    \begin{tabular}{c c p{4cm}p{6cm}}
        \toprule
        \textbf{PID}& \textbf{Gender} & \textbf{Job Role} & \textbf{Work Experience in Data Science}\\
        \midrule
        P1 & M & Expert Data Scientist & 5-10 years \\
        P2 & M & Application Developer & 3-5 years \\
        P3 & M & Novice Data Scientist & less than 3 years \\
        P4 & M & Novice Data Scientist & 0 year (just start learning data science) \\
        P5 & M & AI Operator or ML Engineer & 3-5 years \\
        P6 & M & Novice Data Scientist & less than 3 years \\
        P7 & M & Application Developer & 3-5 years \\
        P8 & M & Novice Data Scientist & less than 3 years \\
        P9 & F & Expert Data Scientist & 3-5 years \\
        P10 & M & Expert Data Scientist & 5-10 years \\
        P11 & M & Expert Data Scientist & more than 10 years \\
        P12 & F & Novice Data Scientist & 3-5 years \\
        P13 & F & Expert Data Scientist & 5-10 years \\
        P14 & M & Novice Data Scientist & 0 year (just start learning data science) \\
        P15 & M & Expert Data Scientist & more than 10 years \\
        P16 & M & AI Operator or ML Engineer & 3-5 years \\
        P17 & M & Subject Matter Expert & 3-5 years \\
        P18 & M & Expert Data Scientist & more than 10 years \\
        P19 & F & Application Developer & 3-5 years \\
        P20 & F & Expert Data Scientist & 3-5 years \\
        P21 & M & Novice Data Scientist & 3-5 years \\
        P22 & M & Novice Data Scientist & less than 3 years \\
        P23 & M & Application Developer & less than 3 years \\
        P24 & M & Novice Data Scientist & less than 3 years \\
        \bottomrule
    \end{tabular}
\end{table}

\subsection{Study Protocol}
We conducted a within-subject controlled experiment with 24 data scientist participants.
Their task was to add documentation to a given draft notebook, which only has code and no documentation at all. The participants were told that they were adding documentation for the purpose of sharing those documented notebooks as tutorials for data science students who just started learning data science.
Each participant is asked to finish two sessions, one with the \sys{} support (Experiment condition) and one without its support (Control Condition). 
We prepared two draft notebooks, one for each session, shown in the Appendix. 
\revision{The two experiment notebooks are adapted from winning notebooks from two Kaggle challenges, which are not included in the model training dataset: 1) House Price Prediction\footnote{\url{https://www.kaggle.com/c/house-prices-advanced-regression-techniques}}; 2) COVID Case Prediction\footnote{\url{https://www.kaggle.com/c/covid19-global-forecasting-week-1}}.
The two notebooks have the same length (9 code cells) and a similar level of difficulties.
Although the two notebooks are simplified versions from winning notebooks, they cover most stages in data science lifecycles.
In addition, the length of the notebooks falls into the middle range of the notebook length distribution on the GitHub corpus (as refer to Figure \ref{fig:rule}).}
To counterbalance the order effect, we randomized the order of the control condition and the experiment condition for each participant, so some participants encountered \sys{} in their first session, some others experienced it in their second session.

Each participant was given up to 12 minutes (720 seconds) to finish one session.
\revision{We conducted three pilot run sessions, and all the three pilot participants were able to finish a single task within 10 minutes, with or without the support of \sys{}.}
Before the experiment condition session, we gave the participant a 1-minute quick demo on the functionality of \sys{}.
All study sessions were conducted remotely via a teleconferencing tool.
We asked the participants to share their screen and we video recorded the entire session with their permission.
After finishing both sessions, we conducted a post-experiment semi-structured interview session to ask about their experience and feedback.
We had a few pre-defined questions such as \say{How do you compare the experience of the documenting task with or without the support of \sys{}?} or \say{Did you notice the multiple candidates in the dropdown menu? Which one do you like the most?} 
In addition, participants were encouraged to tell their stories and experience outside these structured questions.
The interview sections of the video recordings were transcripted into text.

\subsection{Data Collection and Measurements}
We have three data sources: the observational notes and video recording for each session (N = 48), the final notebook artifact out of each session (N = 48), and the post-task questionnaire and interview transcripts (N = 24).

Our first group of measurements are from coding participants' behavioral data from the session recordings. 
In particular, we counted \emph{the task completion time} (in secs) for all sessions. 
Then, for experiment condition only, we also counted the followings: how many times a participant clicked on the light bulb icon to check for suggestions (\emph{code cells checked for suggestions}); how many times a participant directly used the generated documentation (\emph{markdown cells created by \sys{}}); how many times a participant ignored the generated recommendations and manually created the documentation (\emph{markdown cells created by human}); and how many time a markdown cell is co-created by human and \sys{}(\emph{markdown cells co-created by human and \sys{}}).
The result of this analysis is reported in Table \ref{tab:usage} and \ref{tab:performance}.

Secondly, to evaluate the quality of the final notebook artifact, we define our second group of measurements by counting: the \emph{number of added markdown cells}, and the \emph{number of added words} as these two are indicators of the quantity and effort each participant spent in a notebook. \revision{Also, we asked the participants to give a self-reported satisfaction score to each of the two documented notebooks.
We considered that score (-2 to 2) as a self-reported subject feeling of the notebook satisfaction.
In parallel, we asked two experts to rate the notebook-level quality (N = 48) with a 3-dimensional rubric (based on \cite{garousi2013evaluating}) to evaluate the documentation's \emph{readability, accuracy, and informativeness} in a notebook.
We considered these three scores (-2 to 2) as an objective quality of the notebook. }
Readability concerns whether the documentation is in readable English grammar and words, while accuracy concerns how the documentation matches the code content, and informativeness evaluates whether the documentation covers more information units.
Two experts iteratively discussed and evaluated the notebooks until the independent ratings achieved an agreement ($\alpha$ = 0.76, Krippendorff's alpha). 
The result of this analysis is reported in Table \ref{tab:usage}.

For the experiment session only, we conducted a cell-level expert rating (N = 194) using the same approach as in notebook-level expert rating.
Two experts iteratively discussed and evaluated the notebooks until the independent ratings achieved an agreement ($\alpha$ = 0.88, Krippendorff's alpha).
The result of this analysis is reported in Table \ref{tab:performance}.
In addition, we asked the participants to finish a post-experiment survey (5-point Likert Scale, -2 as strongly disagree and 2 as strongly agree, Figure~\ref{fig:questionnaire}) to collect their feedback specific on the system's \emph{usability, accuracy, trust, satisfaction, and adoption propensity} (based on \cite{weisz2019bigbluebot}).

Lastly, for the interview transcripts, four researchers of this research project conduct an iterative open coding method to get the code, theme, and representative quotes as the third group of data.
They each independently coded a subset of interview transcripts, and discussed the codes and themes together. 
After the discussion, they when back and reiterated the coding practice to apply the codes and themes to their assigned notebooks. 
Some examples of the identified themes are:  pros and cons of \sys{}; preference of the three document generation approaches; future adoption, and suggestions for design improvement. 
We will report the qualitative results as supporting materials together with reporting the quantitative results.  

\subsection{Results}

\begin{table}[t]
\caption{\revision{Performance data in two conditions (M: mean, SD: standard deviation): the task completion time (secs), participants' satisfaction with the final notebook (from -2 to 2), graded notebook quality, number of markdown cells, and number of words. In particular, participants spent less time to complete the task in the experimental condition than the control condition (p = .001); participants were more satisfied with the final notebook in the experimental condition than the control condition (p = .04).}} 
\label{tab:usage}
\centering
\footnotesize	
\begin{tabular}{p{5.5cm}cccl}
 \toprule
 &\textbf{Condition} & \textbf{M} & \textbf{SD} &  \\
  \midrule
  \vspace{7pt}
\multirow{2}{*}{Number of Added Markdown Cells} 
& Experiment & 8.04 & 2.40 & 
\multirow{2}{*}{
\includegraphics[width=3.5cm]{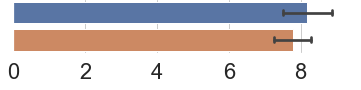}}  \\
& Control & 7.79 & 1.91 & \\
\midrule
  \vspace{7pt}
\multirow{2}{*}{Number of Added Words} 
 & Experiment & 95.75 & 50.56 & 
\multirow{2}{*}{
\includegraphics[width=3.5cm]{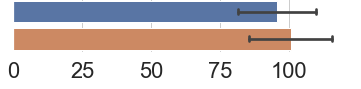}}  \\
 & Control & 100.92 & 53.27 & \\
\midrule
  \vspace{7pt}
\multirow{2}{*}{\textbf{**Task Completion Time (secs)}} 
& Experiment & 391.12 & 200.15 & 
\multirow{2}{*}{
\includegraphics[width=3.5cm]{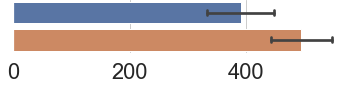}}  \\
& Control & 494.75 & 184.28 & \\
\midrule
  \vspace{7pt}
\multirow{2}{*}{\textbf{*Satisfaction with the Final Notebook (-2 to 2)}} 
& Experiment & 0.96 & 0.69 & 
\multirow{2}{*}{
\hspace{-6pt}
\includegraphics[width=3.5cm]{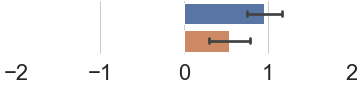}}  \\
& Control & 0.54 & 0.83 & \\
\midrule
  \vspace{7pt}
\multirow{2}{*}{Expert Rating: Accuracy (-2 to 2)} 
& Experiment & 1.60 & 0.47 & 
\multirow{2}{*}{
\hspace{-4pt}
\includegraphics[width=3.5cm]{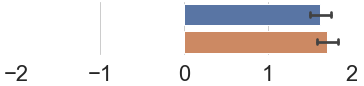}}  \\
& Control & 1.62 & 0.52 & \\
\midrule
  \vspace{7pt}
\multirow{2}{*}{Expert Rating: Readability (-2 to 2)} 
& Experiment &  0.65 & 0.83  & 
\multirow{2}{*}{
\hspace{-4pt}
\includegraphics[width=3.5cm]{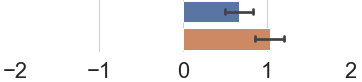}}  \\
& Control & 0.90 & 0.57 &\\
\midrule
  \vspace{7pt}
\multirow{2}{*}{Expert Rating: Informativeness (-2 to 2)} 
& Experiment & 0.67 & 0.64 & 
\multirow{2}{*}{
\hspace{-4pt}
\includegraphics[width=3.5cm]{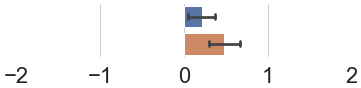}}  \\
& Control  & 0.75 & 0.63 & \\
\bottomrule
\end{tabular}
\end{table}

In this section, we present the user study results on: how \sys{} improved participants' performance on the task, how participants perceived the documentation generation methods in \sys{}, and how participants described the practical applicability of \sys{}.

\subsubsection{\sys{} supports participants to easily add documentations to a notebook. }
Our experiment revealed that \sys{} improved participants' performance on the task by reducing task completion time and improving the satisfaction with the final notebooks.

We performed a two-way repeated measures ANOVA to examine the effect of the two notebooks and the two conditions (with or without \sys{}) on task completion time.
As shown in Table \ref{tab:usage}, participants spent significantly \textbf{less time} (p<.001) to complete the task using \sys{} in the experiment condition (M(SD) = 391.12 (200.15)) than in the control condition (M(SD) = 494.75 (184.28)).
In addition, there was not a statistically significant effect of notebooks on task completion time, nor a statistically significant interaction between the effects of notebooks and conditions on completion time. 

\begin{figure}[t]
    \includegraphics[width=\textwidth]{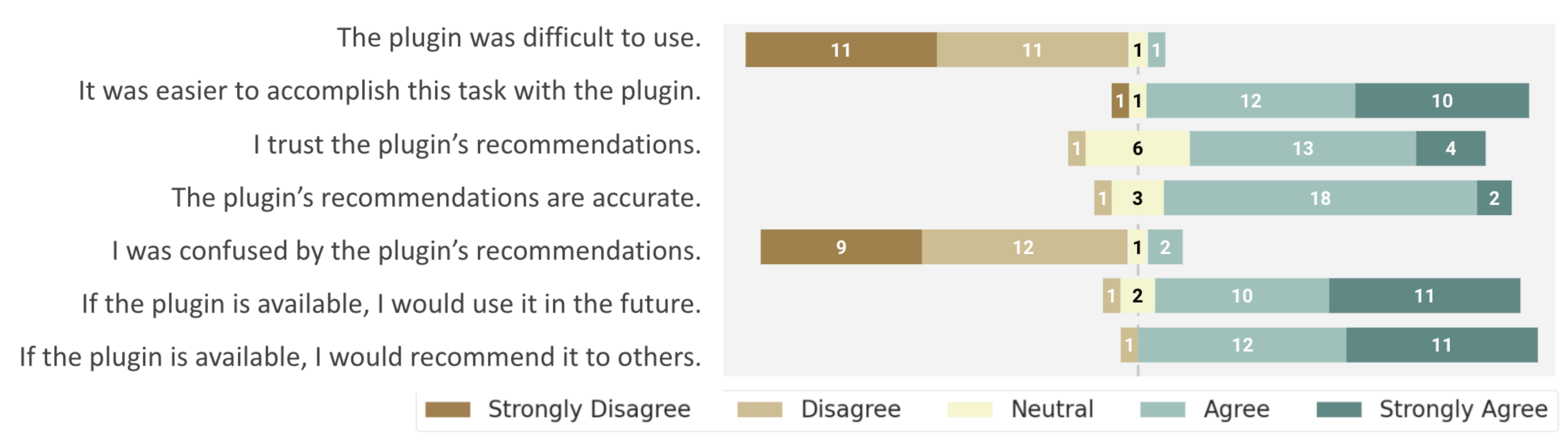}
    \caption{\revision{Results of the post-task questionnaire. Note that the disagrees are not from the same participant.}}
    \label{fig:questionnaire}
\end{figure}

The post-experiment survey result supported our findings. Most participants believed it was easier to accomplish the task with \sys{}'s help (22 out of 24 rated agree or higher), as shown in Figure \ref{fig:questionnaire}. 
And the \sys{} generated recommendation was accurate (20 out of 24 rated agree or higher).

Looking into the qualitative interview data, we can find some potential explanations for why participants believed so. 
Participants reported that \sys{} provided them something to begin with, thus it was easier than starting from scratch: \inlinequote{The plugin makes it easy to just pick it and have something simple. And then I got a couple of times where I went back and said, \say{Oh let me add a few more words.}} (P21).

\subsubsection{Co-creation yields longer documentation and improves accuracy and readability.}
Through coding the video recordings for only the experiment-condition sessions, 
we were able to examine the following questions: while the \sys{} was available, how did the participants use it? Did they check the recommendations it generated? Did they actually use those recommendations in their documentations added into notebooks?

As shown in Table \ref{tab:performance}, we found that while \sys{} is available, for 86.11\% of code cells, participants checked the recommended documentation by clicking on the light bulb icon to show the dropdown menu.
Then, 46.90\% of the created markdown cells were directly adopted from \sys{}'s recommendation; while 11.86\% of the created markdown cells were manually crafted by humans alone. 
\textbf{The most interesting finding is that 41.24\% markdown cells were co-created by \sys{} and human participants together}: \sys{} suggests a markdown cell, human participants take it, and modify on top of it.
This result suggested that most participants used \sys{} in the creation of documentation, and some of them formed a small collaboration between humans and the \ac{AI}.
This finding inspires us to further explore how participants co-create the documentation with \sys{} \cite{muller2021hai}.
By looking at the log data, we discover several editing patterns.
For example, many participants added supplemental details (e.g., expanding the steps into substeps) to Themisto's suggested documentation.
Participants also added stylistic edits, including modifying document hierarchies, polishing sentences, and changing conversational tones.

In order to explore the differences among documentation created by three methods (created by \sys{} only, co-created by human and \sys{}, created by human only), we conducted a cell-level expert rating (N = 194) along the dimension of accuracy, readability, and informativeness.
We also calculated the word count of the documentation length.
We performed a one-way ANOVA to examine the differences among the three groups.
As shown in Table \ref{tab:performance}, markdown cells that are co-created by humans and \sys{} have significantly more word count (M(SD) = 15.45 (10.97)) than markdown cells that are manually written by humans alone (M(SD) = 10.26 (7.41)) and the markdown cells that are directly adopted from \sys{}'s recommandation (M(SD) = 8.88 (7.14)), with F = 11.83, p < 0.001 .
Markdown cells co-created by humans and \sys{} also yield better results in terms of accuracy (F = 9.43, p < 0.001) and readability (F = 3.28, p = 0.04), while for informativeness, there is no significant differences across three groups.
Our posthoc analysis suggested that no significant differences were found between markdown cells created by \sys{} and markdown cells created by humans only along all dimensions (including word count, accuracy, readability, and informativeness).
\begin{table}[t]
\caption{Usage data of the plugin in experimental condition. The results indicate that participants used the plugin for recommended documentation on most code cells (86.11\%). For markdown cells in the final notebooks, 46.90\% were directly adopted from the plugin's recommendation, while 41.24\% were modified from the plugin's recommendation and 11.86\% were created by participants from scratch.}
\label{tab:performance}
\scriptsize
\begin{tabular}{p{4.5cm}p{0.2cm}p{0.5cm}p{1.6cm}p{1.25cm}p{1.2cm}p{1.8cm}} 
 \toprule
 & \textbf{$N$}  & \textbf{$\%$}  & \textbf{Word Count} & \textbf{Accuracy} & \textbf{Readability} & \textbf{Informativeness} \\
 \midrule
    Code Cells Checked for Suggestions & 186 & 86.11 
    & - & - & - & - \\
    Markdown Cells Created by \sys{} Only & 91 & 46.90 
    & 8.88 (7.14) & 1.36 (0.54) &  1.94 (0.23)& 1.34 (0.56)  \\
    Markdown Cells Co-created by Humans and \sys{} & 80 & 41.24 
    & 15.45  (10.97) & 1.68 (0.47) & 1.96 (0.17) & 1.51 (0.55)\\
    Markdown Cells Created by Humans Only & 23 & 11.86 
    & 10.26 (7.41) & 1.28 (0.69) & 1.83 (0.36) & 1.35 (0.65)\\
 \bottomrule
\end{tabular}
\end{table}

\subsubsection{\sys{} increases participant's satisfaction , while maintaining a similar quality of the final notebook.}
The post-task questionnaire revealed that participants were \textbf{more satisfied} with the final notebook after using \sys{} in the experiment condition than in the control condition (p = .04) (Table \ref{tab:usage}). 
The interview results also supported this finding. 
P14 believed that \sys{} helped with wording: \inlinequote{Sometimes I knew what the cells were doing but I did not know how to put things in a really good sentence for others.}

\sys{} also motivates participants to document the analysis details.
Although we did not see a difference in the number of markdown cells created in two conditions or the number of words in total, \sys{} helps them overcome the procrastination of writing documentation and reminds them to document things that they might ignore.
\begin{quote}
    I think I definitely overlooked some details when I was commenting without the tool, because I just made the assumption that people should know from the code... To be honest, I do not usually follow a good coding practice. My notebooks are really messy and I am the only person who can understand it. I feel sorry for anybody else that has to see it. (P19)
\end{quote}

Moreover, participants believed that \sys{} can help them form a better documenting practice in the long term: \inlinequote{It very useful to remind me to always put some documentation in a timely manner.} (P13).

The two experts' gradings for the notebook quality suggest that there was not a significant difference for the three dimensions of the quality rubric (\textbf{accuracy, readability, and completeness}).
In the post-task interview, participants mentioned that the accuracy of the generated recommendations plays a role in participant's experience: \inlinequote{My experience with the plugin is definitely better. For the most part, the suggestions are pretty accurate. Although sometimes I did make a few minor changes like rearranging the text.} (P5).
Some participants also mentioned that they needed to edit the format of the generated document to fit their context.
We believe that while \sys{} offers convenience to improve the data scientists' productivity and saves their time, it may not provide the same level of readability as those notebooks well articulated by humans. 
Thus, data scientists may want to further revise the formatting and wording of the \sys{} generated documentation.

In summary, our experiment indicated that \sys{} improves participants' productivity for creating documentation. 
It also increases their perceived satisfaction with the final notebook, compared to the notebooks written by participants themselves.

\subsubsection{The three approaches of generating documentation are suitable for different scenarios.}
In this section, we have an in-depth analysis of how participants perceived the three different approaches that \sys{} implemented to generate documentations: the deep-learning-based approach, the query-based approach, and the prompt-based approach. 
In the post-experiment interview, we explained how \sys{} generated the documentation with these three approaches, and asked participants if they like or dislike one particular approach.

Participants reported that they felt the deep-learning-based approach provided concise and general descriptions of the analysis process: \inlinequote{I think the AI suggestion gives me an overview. It is short, and has some useful keywords.} (P12).

Participants also suggested that the deep-learning-based approach sometimes generated inaccurate or very vague documentation: \inlinequote{The first one gives me a very short summary, though it didn't always say what the cell is doing.} (P1).
But the deep-learning-based approach is still perceived useful. 
As it is short and with only a couple of keywords, many participants believe it may be more suitable for some quick and simple documentation task, or for the analyst audience who can understand these short keywords.

In terms of the query-based approach, participants believed that the documentation generated from this approach contains has longer and more descriptive information.
This approach is further perceived to be more suitable for educational purposes: \inlinequote{This one gives you really good information. For some specific methods or calls, you don't have to come up with a high-level summary for others and you can directly use it.} (P14).

Participants also acknowledged that such a query-based approach may not work for some scenarios.
For example, participants found that the query-based approach was not useful for summarizing the very fundamental level data manipulations, as there was no core \ac{API} method in it. 
Some participants mentioned that the usefulness of this query-based approach depends on the audience. 
\begin{quote}
    The [deep-learning-based approach] was really useful. The [query-based approach]... it depends on the audience. It is much more appropriate for a novice programmer. (P18)
\end{quote}

We observed in the video recordings that participants rarely used the prompt-based approach in the session.
The interview data confirmed our speculation. 
Some participants said that they liked the idea of user prompts, but they did not use it because the deep-learning-based approach and the query-based approach already gave them the actual content.
Other participants pointed out that the prompts were not intelligent enough, so they did not use it: \inlinequote{It always asks the same thing and I just ignored the prompts.} (P18).

Participants suggested that the prompts could be designed to better fit the context.
\begin{quote}
    Perhaps the system can infer what the code cell was doing [from the deep-learning-based approach], and show prompts accordingly. Like if I delete a data point from the dataset, there is a prompt asking why I considered it as an outlier or something. (P5)
\end{quote}

Last but not the least, many participants preferred a hybrid approach to combine the deep-learning-based approach and the query-based approach.
For example, P12 mentioned,
\begin{quote}
    The first one (deep-learning-based) tells me what the code cell is doing in general and the second one (query-based) tells me the details of the function. I would go with a hybrid approach. (P12)
\end{quote}

\subsubsection{Will participants use \sys{} in their future data science project?}
Most participants indicated that they would like to use \sys{} in the future when answering the survey question as shown in Figure \ref{fig:questionnaire}.
The interview data provides more detail and evidence to elaborate on this result. 
Participants suggested various scenarios in which the \sys{} could be useful in their future work, such as they need to add documentation during the exploration process for future selves, or they need to document a notebook in a post-hoc way for sharing it with collaborators, or they need to mentor a team member who is a novice data scientist, or they need to refactor an ill-documented notebook written by others:
\begin{quote}
    When I am doing data analysis, I tend to write the code first because there is a flow in my head of what I need to do. And then I will go back afterward to use the plugin and add the comments needed. I will definitely do that before sharing that file or handing it over to others. (P12)
\end{quote}

There was one participant who did not think \sys{} could fit into his workflow: \inlinequote{I always write documentation before writing code. Maybe \sys{} does not work for people like me.} (P3)

In our experiment, we provided the scenario as they were documenting the notebook as a tutorial for some data science students. 
In the interview, we asked how participants would document a notebook differently if they were created documentation for the notebooks for non-technical domain experts audience.
Some participants suggested that computational notebooks may not be a good medium to present the analysis to non-technical domain experts.
They would prefer to curate all the textual annotations in a standalone report or slide decks.
Some others believed that the notebook could work as the medium but they would change the documentation by using less technical terminology, adding more details on topics that the non-technical domain experts would be interested in (e.g., how data is collected, potential bias in the analysis).

\subsubsection{Participants suggest various design implications for automated code documentation.} 
In the interview, participants provided various design suggestions to improve \sys{} and to design future technologies that can support data scientists to document the notebook.

Participants expected \sys{} to have more functionalities than simply generating documentation for the code. 
For example, P13 proposed maybe \sys{} can also create a description to document the changes of versions and the editing histories from different team members. 
P3 and P4 believed that the automatic generation of Reason is very much needed for explaining decisions such as why selecting a particular algorithm. 
P19 wanted the system to automatically add explanations to the execution errors.
Participants also mentioned that \sys{} should add more varieties into the generated content's formatting.
They would like to see suggested documentation with a better presentation.

And lastly, some participants suggested that maybe such a documentation generation system can take consideration of the purpose of the notebook, the domain-specific terminology, or the indivials' habits for writing documentation.

\subsubsection{Summary of the Results}
In summary, our study found that \sys{} can support data scientists in generating documentation by significantly reducing their time spent on the task, and improving the perceived satisfaction level of the final notebook.
When \sys{} is available, participants are very likely to check the generated documentation as a reference. 
Many of them directly used the generated documentation, a few of them still prefer to manually type the documentation, while many of them adopted a human-AI co-creation approach that they used the AI-generated one as a baseline and keep improving on top of it.
Participants perceived the documentation generated by the deep-learning-based approach as a short and concise overview,  the documentation generated by the query-based approach as descriptive and useful for educational purposes, but they rarely used the prompt-based approach.
Overall, participants enjoyed \sys{} and would like to use it in the future for various documenting purposes.

\section{Discussion}
\subsection{The Documentation Practices in Data Science is Different from in Software Engineering}
The practice of documentation in data science has both overlaps and strong contrasts in relation to the ones in software engineering in many facets.
Software engineers write inline comments in their work-in-progress code to help collaborators understand the behavior of the code without the burden of going through thousands of lines of code;
they document changes of their code for better version management and improving awareness of their collaborators;
when others need to build upon their work, they write formal documentation and Readme files to describe how to use functions and API in their packages or services~\cite{lowndes2017our,saltz2017comparing,zhang2020data}.
Data scientists write computational narratives as a practice of literate programming~\cite{kery_exploring_2017, Park:2018:PPL:3266037.3266098}, and as a way to think and explore alternatives. 
Thus, notebooks often have orphan code cells or out-of-order code snippets, which leads to lower reusability of the notebook and further highlights the importance of documentation in the notebook. 
As we found in our formative study, well-documented notebooks explain more than the behavior of the code.
Notebooks cover various topics including describing and interpreting the output of the code, explaining reasons for choosing certain algorithms or models, educating the audience from different levels of expertise, and so on.

Thus, many interventions and lessons learned about documentation in software engineering may not apply in data science context.
For example, how can we evaluate the quality of the documentation?
Software documentation can be assessed based on attributes like completeness, organization, the relevance of content, readability, and accuracy~\cite{garousi2013evaluating}.
Our experiment found that the quality scores assessed by these rubrics does not reflect users' satisfaction with the final notebooks.
Despite many people's efforts to creating a standard documentation practice \cite{rule2019ten, konkol2020publishing}, it remains questionable whether there is a one-size-fits-all solution.
For example, Rule et al. \cite{rule2019ten} suggested ten rules for writing and sharing computational analysis in Jupyter notebooks.
The first rule they proposed is to tell a story to the audience.
However, this description is very general as people may approach storytelling differently.
As we observed in Kaggle notebooks, some notebook authors prefer to use concise and accurate language while others use more colloquial and creative language.
These creative notebooks stand out and receive many votes and compliments from the Kaggle community.
As we recognize documentation in data science as a fluid activity, traditional template-based approaches to aid documentation writing may not work in data science because they can not capture a broader aspects of documentation, and limit the expressiveness of storytelling.

We argue that future work should recognize the difference between data science and software engineering, and tailor the documentation experience for data scientists. 
For example, Callisto \cite{wang2020callisto} harnessed the fact that data scientists engage in synchronous work and discussion, and used contextual links between discussion messages and notebook content to aid the explanation of notebooks.

\subsection{Human-AI Collaboration in Code Documentation in Data Science}

\revision{We argue that AI-assisted code documentation process can be viewed as a co-creative process in which machine learning fits into the human workflow and collaborate together to create documents in a notebook. }
The notion of a ``partnership relationship'' between human data scientists and AI has been discussion by Wang et al.~\cite{wang2019human}, and is part of a larger research discussion by many others (e.g.,~\cite{cai2019human,seeber2020machines,malone2018human}).
\revision{We consider this partnership as broadly defined where an AI system does not need an avartar or a conversational interface, but this AI system should be designed to fit into the existing human workflow and assist some parts of the human task to improve the quality or productivity.
Human-AI collaboration, as opposed to human-AI competition (portraited by AlphaGo or DeepBlue), should be the ultimate goal of human-AI interaction research.
Various human-centered AI design principles (such as human-in-the-loop) are means to get to this end goal.
Our study demonstrated another means to achieve human-AI collaboration, where we combined the fully automated neural network approach and the less advanced rule-based or prompt-based approaches.
This design is to acknowledge the limitation of today's neural-network modeling.}
Our result showed that the combined human + AI effort produced a satisfied level of  quality at a \emph{more rapid pace} than what human or AI could achieve alone. 

We also observed the user interaction pattern in which the AI creates \say{first draft} of the documentation, followed by human review and editing, resulted in a final artifact that not only met the bar for quality, but exceeded it for the level of satisfaction. 
Participants were happier with their code documentation when they were assisted by the AI system to create it, rather than when they worked alone. 
Thus, we conclude that the benefits of having an AI partner in this task stem from being able to produce the same high level of objective quality, but at a much more rapid pace (20\% faster on average) and with a higher level of satisfaction with the end product.

We speculate that one of the contributing factors for why people were accepting of the AI's suggestions is because the final decision of taking those suggestions was up to the human. 
\revision{As an alternative, we could have designed the system to always automatically produce a markdown documentation cell for each identified code cell, but we decided not to.
Because this fully-automated design is an extreme in the framework of automation put forth by Parasuraman et al.~(\cite{parasuraman2000model}; see also \cite{horvitz1999principles}), which people may feel being replaced. 
Our results confirmed our assumption --- }participants reported that they enjoyed being able to see multiple suggestions, created using different algorithms, and select the one that was the closest match to their intent in documenting a code cell. 
This level of interaction corresponds to \say{AI executes a selection only after a human has approved} in the Parasuraman et al. model \cite{parasuraman2000model}. 

Our result also shed light on the research question in~\cite{wang2019human} about the conditions under which human data scientists will enjoy working with AI partnership . 
In our case, maintaining control of the initiative and the final decision is an important aspect for people's enjoyment and acceptance of the AI system. 
It remains to be studied whether people prefer both to control their own initiative \textit{and} the initiative of a machine teammate, as proposed in Shneiderman's recent two-dimensional model~\cite{shneiderman2020human}.
\revision{Also, we did not focus on the explanability or trust aspect of the designed AI system, such as how to visualize the connection between the generated documentaiton and the original code.
In the future, the explanability and trust aspects of the AI system in the data science context is a very critical research topic (e.g., ~\cite{weidele2020autoaiviz,drozdal2020trust}), and should also be prioritized in the research agenda.}

\revision{There are many other tasks in a data science project's lifecycle that could use AI's help, such as model presentation or feature engineering~\cite{automationsurvey}. In the future, we plan to extend our work to design more human-centered AI systems to support users in these data science tasks as well.}

\revision{In the future, we plan to explore whether the identified benefits and tradeoffs persist or not after a long period of adoption by users.
One of the potential benefits could be: the human-AI collaboration work style helps users to learn more from the AI suggested/reminded documentations, thus they realize more of the value of adding code documentation to notebooks; in contrary, maybe users become over-reliance on AI systems thus they de-skill in this code documentation task, both hypotheses await future research to evaluate.}

\subsection{Design Implications}
We offer designers and tool builders the following suggestions to encourage data scientists to write better documentation:
\subsubsection{Towards Hybrid and Adapted Code Summarization}
Our evaluation of \sys{} indicates that instead of a fully automatic approach, data scientists prefer to use a hybrid method for helping them write documentation.
We argue that future work for code summarization should investigate a hybrid and adapted approach.
We suggest that \textit{adaptive interactive prompting} may be a worthwhile research topic.
For example, prompts could be based on the contents of the code cell which the user was trying to document. 
Another possibility is that prompts be based on the user's own history of writing markdown cells, and could either appeal to the user's strengths, or could anticipate and accommodate the user's weaknesses. 
In a more socially-oriented approach, users within an organization might rate the initial set of prompts, voting some prompts up or down depending on their usefulness. An evolution of this idea might allow users to propose new prompts for use by selves and others (e.g., \cite{dimicco2009people}).
Furthermore, we argue that future code summarization tools would benefit from a reinforcement learning approach which learns from users' modifications to the original proposed texts, and could anticipate the users' preference in subsequent documentation. 

\subsubsection{Customizing the Recommendations based on Usage Scenarios} 
As prior work stated \cite{zhang2020data}, data science workers engage in various collaborations during different stages of the data science lifecycle. Documentation plays an important role in many scenarios.
For example, handing off work between data engineers, communicating results with stakeholders, or informal notes to future self.
Data science workers may have different needs of the documentation for different usage scenarios.
Designers and tool builders should take a user-centered approach to understand the purpose of documentation, the appropriate level of details, and the best way to present the documentation.
For example, participants suggested that future versions of \sys{} being able to document the changes of versions, errors, and related online forum posts.
Participants also suggested that they would like to see more varieties into the generated narrative's formatting. 

\subsubsection{Inverting \sys{} -- Automatic Code Generation from Documentation.}
The premise of \sys{} was to generate descriptive material based on program code.
Following some of the ideas in Seeber et al.~\cite{seeber2020machines}, we might invert this strategy. 
We recall that P3 told us that he wrote documentation in advance of writing the code itself. 
If there are other people who use the same discipline as P3, could we generate code from descriptive text? 
We suspect that this idea would not work for just \textit{any} textual description. 
However, there could be certain stylized ways of writing descriptions that might be translatable into code; pseudocode could provide a starting point for the design of such a stylized type of description. 
We recognize that this kind of approach would need to have a representation of code packages and libraries, so that it could generate code that was appropriately structured for those packages. 
Of course, package documentation could be used to construct such a representation.

\subsection{Limitation}
\revision{Our formative study only explores notebooks from the Kaggle corpus, which may leave out some varieties of markdown cells that only exist in messy notebooks that can benefit from the support of documentation generation.
However, notebooks on the Kaggle platforms are based on real data and real problems, and they aim for rich explanations and narratives, where other places do not have high-quality notebooks with rich documentation. 
Future work should expand the exploration on the other notebook corpus, for example, notebooks published with scientific paper which contain fine-grained documentation.}

Our experiment has several limitations: it focuses only on the documenting (instead of coding) process, it is a controled experiment study, and participants did not work on notebooks created by themselves.
Thus, for example, we do not know how participants would perceive the usefulness of the tool in \revision{realistic notebooks}, which may be longer and more complicated (e.g., having out-of-order cells) than the notebooks we provided.
However, we believe the result is still promising to shed light for future research and future system design.
Future work can explore the generalizability of \sys{} through a long-term deployment study.

As for the Human-AI Collaboration research initiative, our work only reports the findings on how human and AI collaborated at a coarse-grained level (Table~\ref{tab:performance}). 
In the future work, we will have an indepth analysis to break down to the level of individual cells, and further analyze the difference between automatically-generated, co-edited, or manually-produced cells. 
This detailed analysis will help us to understand how human behave and perceive the fine-grained collaboraiton and interaction with the AI partner. 
And such findings and its derived design insights could also help researchers who are studying Human-AI Collaborations in other usage scenarios (e.g., in Healthcare or in Educational settings~\cite{xu161same}) beyond the nodebook documentation context in this paper.

\section{Conclusion}
In this paper, we have designed and built \sys{} to support human data scientists in the notebook documentation task.
This researhc prototype also serves as a prompt to explore the human-AI collaboration research agenda within the automated notebook documentation user scenario.
The system design is driven by insights from previous literature, and also by a formative study that analyzed 80 highly-voted Kaggle notebooks to understand how human data scientists document notebooks.
The follow-up user evaluation suggested that the collaboration between data scientists and \sys{} significantly reduced task completion time and resulted in a final artifact that not only met the bar of quality, but also exceed it for the level of satisfaction.


 \appendix
\section{Example of Documentation Generation in Themisto}
 \begin{landscape}
\begin{table}[h]
\footnotesize
\centering
\caption{\revision{Example notebook (House Prediction), documentation generation in Themisto, and user submitted documentation (T - Markdown Cells Created by Themisto Only, C - Markdown Ceells Co-created by Humans and Themisto, H - Markdown Cells Created by Humans Only).}}
\label{tab:notebook-house}
\begin{tabular}{p{5.5cm}p{2cm}p{3.5cm}p{2cm}p{3.8cm}} 
 \toprule
 \textbf{Source Code} & \textbf{DL-Based} & \textbf{Query-Based} & \textbf{Prompt-Based} & \textbf{P2}\\
 \midrule
\begin{tabminted}{python}
import pandas as pd
import numpy as np
from sklearn.linear_model import LassoCV
from sklearn.model_selection 
    import cross_val_score
\end{tabminted}
& Importing libraries 
& Pandas is for data manipulation and analysis; NumPy is a library for ... 
& This code cell is for \_ \_ \_ \_ \_ 
& Importing libraries (T)\\
\midrule
\begin{tabminted}{python}
train = pd.read_csv('train.csv')
test = pd.read_csv('test.csv')    
\end{tabminted}
& Read the data
& Read a comma-separated values (csv) file into DataFrame; Return the first 5 rows.
& This code cell is for \_ \_ \_ \_ \_ 
& Read the data (T)\\
\midrule
\begin{tabminted}{python}
train.head()
\end{tabminted}
& Let's see the values
& Return the first 5 rows
& The table shows \_ \_ \_ \_ \_ 
& Return the first 5 rows. (defValue=5) (C)\\
\midrule
\begin{tabminted}{python}
all_data = pd.concat((
train.loc[:,'SubClass':'SaleCond'],
test.loc[:,'SubClass':'SaleCond'])) 
\end{tabminted}
& A generator for feature
& Concatenate pandas objects along a particular axis with optional set logic along the other axes.
& This code cell is for \_ \_ \_ \_ \_ 
& Concat train and test col "SaleCondition" (C)\\
\midrule
\begin{tabminted}{python}
all_data = pd.get_dummies(all_data)
\end{tabminted}
& Convert all the data
& Convert categorical variable into dummy/indicator variables
& This code cell is for \_ \_ \_ \_ \_ 
& Convert categorical variable into dummy/indicator variables. (T)\\
\midrule
\begin{tabminted}{python}
all_data = all_data
    .fillna(all_data.mean())
 \end{tabminted}
 & Check the missing values
 & Fill NA/NaN values using the specified method
 & This code cell is for \_ \_ \_ \_ \_ 
 & \\
 \midrule
\begin{tabminted}{python}
X_train = all_data[:train.shape[0]]
X_test = all_data[train.shape[0]:]
y = train.SalePrice
 \end{tabminted}
 & Create the target and the test data
 & Slice string
 & This code cell is for \_ \_ \_ \_ \_ 
 & \\
 \midrule
\begin{tabminted}{python}
model_lasso = LassoCV(
    alphas = [1, 0.1, 0.001, 0.0005])
    .fit(X_train, y)
 \end{tabminted}
 & Model
 & Lasso linear model with iterative fitting along a regularization path.
 & This code cell is for \_ \_ \_ \_ \_ 
 & Fit regression model (H)\\
 \midrule
\begin{tabminted}{python}
def rmse_cv(model):
    rmse= np.sqrt(-cross_val_score(
        model, X_train, y, scoring=
        "neg_mean_squared_error", cv = 5))
    return(rmse)
rmse_cv(model_lasso).mean()
 \end{tabminted}
 & A simple example model with the lasso
 & Evaluate a score by cross-validation
 & The result indicates that \_ \_ \_ \_ \_ 
 & Define score function and evaluate (H)\\
 \bottomrule
\end{tabular}
\end{table}
\end{landscape}

\begin{landscape}
\begin{table}[ht!]
\footnotesize
\centering
\caption{\revision{Example notebook (Covid Prediction), documentation generation in Themisto, and user submitted documentation (T - Markdown Cells Created by Themisto Only, C - Markdown Ceells Co-created by Humans and Themisto, H - Markdown Cells Created by Humans Only).}}
\label{tab:notebook-covid}
\begin{tabular}{p{5.5cm}p{2cm}p{3.5cm}p{2cm}p{3.8cm}} 
 \toprule
 \textbf{Source Code} & \textbf{DL-Based} & \textbf{Query-Based} & \textbf{Prompt-Based} & \textbf{P5}\\
 \midrule
\begin{tabminted}{python}
import numpy as np
import pandas as pd
from sklearn.ensemble 
    import RandomForestClassifier
\end{tabminted}
& Importing libraries 
& Pandas is for data manipulation and analysis; NumPy is a library for ... 
& This code cell is for \_ \_ \_ \_ \_ 
& Importing libraries (T)\\
\midrule
\begin{tabminted}{python}
train = pd.read_csv("train.csv")
test = pd.read_csv("test.csv")
train.head()    
\end{tabminted}
& Read the data
& Read a comma-separated values (csv) file into DataFrame; Return the first 5 rows.
& The table shows \_ \_ \_ \_ \_ 
& Read and sanity check the data (C)\\
\midrule
\begin{tabminted}{python}
train.describe()
\end{tabminted}
& Let's see the values
& Generate descriptive statistics. Descriptive statistics include ...
& The table shows \_ \_ \_ \_ \_ 
& \\
\midrule
\begin{tabminted}{python}
train["Date"] = train["Date"]
    .apply(lambda x: x.replace("-",""))
train["Date"] = train["Date"]
    .astype(int)
train.head() 
\end{tabminted}
& Convert all the data
& Replace a specified phrase with another specified phrase
& The table shows \_ \_ \_ \_ \_ 
& Preprocess the data (C)\\
\midrule
\begin{tabminted}{python}
train.isnull().sum()
\end{tabminted}
& Check the missing values
& Detect missing values for an array-like object
& The result indicates that \_ \_ \_ \_ \_ 
& Check the missing values (T)\\
\midrule
\begin{tabminted}{python}
test["Date"] = test["Date"]
    .apply(lambda x: x.replace("-",""))
test["Date"]  = test["Date"]
    .astype(int) \end{tabminted}
& Convert all the data
& Replace a specified phrase with another specified phrase
 & This code cell is for \_ \_ \_ \_ \_ 
 & Preprocess the date column (C)\\
 \midrule
\begin{tabminted}{python}
x = train[['Lat', 'Long', 'Date']]
y = train[['ConfirmedCases']]
x_test = test[['Lat', 'Long', 'Date']]
\end{tabminted}
 & Create the target and the test data
 & Select subsets of data
 & This code cell is for \_ \_ \_ \_ \_ 
 & Create the train/test data and the target (C)\\
 \midrule
\begin{tabminted}{python}
Tree_model = RandomForestClassifier(
    max_depth=200, 
    random_state=0)
Tree_model.fit(x,y)
 \end{tabminted}
 & Model
 & A random forest is a meta estimator that fits a number of decision tree classifiers on ...
 & This code cell is for \_ \_ \_ \_ \_ 
  & Define and configure the model

A random forest is a meta ... We also train the model with `.fit()` (C)\\
 \midrule
\begin{tabminted}{python}
pred = Tree_model.predict(x_test)
pred = pd.DataFrame(pred)
pred.columns = 
    ["ConfirmedCases_prediction"]
 \end{tabminted}
 & Predicate to use a predicate function for tests
 & A random forest is a meta estimator that fits a number of decision tree classifiers on ...
 & This code cell is for \_ \_ \_ \_ \_ 
 & Run the model to generate predictions on the test data and store them as a `DataFrame`(H) \\
 \bottomrule
\vspace{1mm}
\end{tabular}
\end{table}
\end{landscape}

\pagebreak
\section{Coding Book for the Interview Transcripts}
\begin{table}[h!]
    \centering
    \caption{Coding Book for the Interview Transcripts}
    \label{tab:codebook}
    \small
    \begin{tabular}{lp{6.5cm}}
    \toprule
         \textbf{Theme} & \textbf{Code}  \\
         \midrule
         \multirow{8}{*}{Pros of Themisto} 
 & Easy to Use \\
 & Provie Inspirations \\
 & Improve Content \\
 & Efficiency \\
 & Hybrid Approach \\
 & Useful for Long Term \\
 & Prefer the Plugin \\
 \midrule
 \multirow{2}{*}{Cons of Themisto}
 & Inaccurate \\
 & Not Useful \\
 \midrule
 \multirow{6}{*}{Perceptions of the Deep-Learning-Based Approach}
 & Concise \\
 & Useful \\
 & Accurate \\
 & Inaccurate \\
 & For Own Use \\
 & For Collaboration Use \\
 \midrule
 \multirow{5}{*}{Perceptions of the Query-Based Approach}
 & Descriptive \\
 & Too Long \\
 & Useful \\
 & Confusing \\
 & Instructive \\
 \midrule
 \multirow{3}{*}{Perceptions of the Prompt-Based Approach}
 & Tedious \\
 & Easy to Use \\
 & Inspiring \\
 \midrule
 \multirow{3}{*}{Future Adoption}
 & Positive Adoption Propensity \\
 & Scenarios for Future Adoption \\
 & Negative Adoption Propensity \\
 \midrule
 \multirow{6}{*}{Design Improvements}
 & More Options Generated by AI \\
 & Handle Presentation and Formatting \\
 & Summarize Other Information (e.g., Reasons, Summary, Errors) \\
 & Custimization \\
 & Optimize UI \\
 & Adaptive Prompts \\
 \bottomrule
    \end{tabular}
\end{table}
\begin{acks}
We thank all of our participants for their help in the study, and the annonymous reviewers for their valuable feedback. 
\end{acks}
\bibliographystyle{ACM-Reference-Format}
\bibliography{refs}










\end{document}